\documentclass[aps,prb,twocolumn,superscriptaddress]{revtex4-2}
\usepackage{graphicx}
\usepackage{amsmath}
\usepackage{amssymb}
\usepackage{mathtools}
\usepackage{xcolor}
\usepackage{braket}
\usepackage{soul}
\usepackage{booktabs} 
\usepackage[normalem]{ulem}
\usepackage{esint}  % used for slash through intgeral
\usepackage[colorlinks=true, citecolor=blue, urlcolor=blue, linkcolor=blue,bookmarks=false,hypertexnames=true]{hyperref} 
\usepackage{appendix}
\usepackage{array}
\usepackage[column=O]{cellspace}
\newcolumntype{P}[1]{>{\centering\arraybackslash}p{#1}}

\newcounter{berryfoot}
\begin{document}
\graphicspath{}
\preprint{APS/123-QED}

\title{Linear response of the Chern insulator MnBi$_2$Te$_4$: A Wannier function  approach}
\date{\today}

\author{Matthew Albert}
\email{matthew.albert@mail.utoronto.ca}
\affiliation{Department of Physics, University of Toronto,
Toronto, Ontario  M5S 1A7, Canada}

\author{Javier Sivianes}
\email{javier.sivianes@ehu.eus}
\affiliation{Centro de F\'{\i}sica de Materiales (CSIC-UPV/EHU), 20018, Donostia-San Sebasti\'an, Spain}

\author{Jason G. Kattan}
\email{jason.kattan@mcgill.ca}
\affiliation{Department of Physics, University of Toronto,
Toronto, Ontario  M5S 1A7, Canada}
\affiliation{Department of Physics, McGill University, Montr\'eal, Qu\'ebec H3A 2T8,  Canada}

\author{Julen Iba\~nez-Azpiroz}
\email{julen.ibanez@ehu.eus}
\affiliation{Centro de F\'{\i}sica de Materiales (CSIC-UPV/EHU), 20018, Donostia-San Sebasti\'an, Spain}
\affiliation{IKERBASQUE, Basque Foundation for Science, 48009 Bilbao, Spain}
\affiliation{Donostia International Physics Center (DIPC), 20018 Donostia-San Sebasti\'an, Spain}

\author{J. E. Sipe}
\email{john.sipe@utoronto.ca}
\affiliation{Department of Physics, University of Toronto,
Toronto, Ontario  M5S 1A7, Canada}

%%%%%%%%%%%%%%%%%%%%%%%%%%%%%%%%%%%%%%%%% Abstract %%%%%%%%%%%%%%%%%%%%%%%%%%%%%%%%%%%%%%%

\begin{abstract}
Recent work demonstrated that in the long wavelength limit the linear response of a Chern insulator to finite-frequency electric fields is the sum of two terms: A general frequency-dependent Kubo contribution that is present irrespective of band topology, and a topological Hall term that vanishes for topologically trivial insulators. Motivated by recent experiments and theoretical predictions, we use these expressions to calculate the optical conductivity and susceptibility of intrinsically magnetic MnBi$_2$Te$_4$ thin films with one, four, five, and eleven septuple layers by combining density functional theory with ``single-shot" Wannier functions. To characterize the underlying topology of these systems, we compute the two-dimensional Chern number of these films using recently derived global expressions formulated in terms of Bloch energies and velocity matrix elements; the use of these expressions allows us to 
circumvent numerical issues at band crossings. Films with eleven septuple layers are of particular interest. We find that they have the same Chern number as five septuple layer films, in contrast to the reported ``higher Chern-number phase" of these systems in other studies; we discuss a few possible reasons for the discrepancy. A layer-resolved decomposition of the static reactive susceptibility reveals a non-uniform spatial distribution, with interlayer contributions dominating over intralayer contributions and the bulk layers accounting for nearly 70\% of the total response. We also 
identify spin-orbit coupling-driven band inversions as a possible indicator of these topological phases.

\vspace{6mm}

\end{abstract}
\maketitle

%%%%%%%%%%%%%%%%%%%%%%%%%%%%%%%%%%%% Introduction %%%%%%%%%%%%%%%%%%%%%%%%%%%%%%%%%%%%%%%%

%%%%%%%%%%%%%%%%%%%%%%%%%%%%%%%%%%%%%%%%%%%%%%%%%%%%%%%%%%
\section{Introduction}
%%%%%%%%%%%%%%%%%%%%%%%%%%%%%%%%%%%%%%%%%%%%%%%%%%%%%%%%%%
The discovery of the quantum Hall effect, in which a system immersed in a uniform magnetic field and subject to a perpendicular applied electric field can exhibit a redistribution of charges described by a \textit{quantized} Hall conductivity %when a perpendicular electric field is applied 
\cite{PhysRevLett.45.494}, was central in the emergence of the new field of ``quantum materials."
From a semiclassical perspective, the background magnetic field induces a kind of ``quantized" cyclotron motion of the electrons therein, and the electric field induces a drift of these cyclotron orbits in the direction orthogonal to both of these fields \cite{Jain_2007}. Importantly, this effect relies on the presence of a uniform magnetic field, and vanishes 
when it is removed. However, it has been demonstrated that the same kind of Hall response can persist in certain materials, called Chern insulators, even in the absence of any background magnetic field, as a consequence of the topological properties of the system’s occupied bands 
\cite{PhysRevLett.61.2015}; this is the \textit{quantum anomalous Hall effect}. Remarkably, this
``topological quantization" of the Hall conductivity is even immune to backscattering and other effects of disorder due,
for example, 
to the presence of impurities or dislocations in the crystal lattice \cite{RevModPhys.82.3045}.
This has led some to suggest the possibility of low-power consumption devices that do not rely on high carrier mobility and strong magnetic fields \cite{bosnar_high_2023,li_high-chern-number_2025}. The robustness of topology extends even further: Recent theoretical work has predicted that while surface reconstructions, likely to occur under experimental conditions, modify the size of the exchange gap, still the topological character of the non-reconstructed system remains preserved \cite{PhysRevResearch.7.023024}.

The first materials found to realize a Chern insulator phase were magnetically doped thin films of the topological insulator (TI) (Bi,Sb)$_2$Te$_3$, using
transition metal dopants
such as chromium (Cr) or vanadium (V) \cite{https://doi.org/10.1002/adma.201703062,doi:10.1126/science.1234414,PhysRevLett.113.137201}. The insertion of these dopants leads to a net magnetization across the sample that breaks time-reversal symmetry. Often accompanied by
a band inversion in the bulk, this can lead to a two-dimensional insulating film with a nonzero Chern number: a Chern insulator.
However, challenges persist in the fabrication of these systems, such as the inhomogeneous distribution of the magnetic dopants and the need for a sufficiently large magnetization, which limit the Chern insulator phase of these materials to very limited regimes \cite{vanderbilt2018berry,doi:10.1126/science.aax8156}. These shortcomings have prompted researchers to investigate intrinsic magnetic systems such as MnBi$_2$Te$_4$, which consists of ferromagnetic septuple layers (SLs) arising from the strong magnetic moments of the Mn atoms. With
alternating magnetization between adjacent SLs \cite{PhysRevResearch.7.023024}, they provide
an ideal platform that naturally combines magnetism and topology.

The quantum anomalous Hall effect was experimentally observed in five SL thin films of MnBi$_2$Te$_4$ \cite{doi:10.1126/science.aax8156}, and it is believed that any thin films containing an odd number of layers (greater than one) will be in a Chern insulator phase \cite{PhysRevLett.122.107202}. In contrast, thin films containing an even number of layers have antiferromagnetic moments from each layer that compensate each other, thus preserving parity-time symmetry and giving rise to a $\mathbb{Z}_2$ topological insulator (sometimes called an ``axion insulator" \cite{PhysRevB.108.125424}) phase, the
leading candidate for the realization of
the topological magneto-electric effect \cite{liu2020robust,PhysRevLett.122.206401,otrokov2019prediction}. It was recently
suggested that the antiferromagnetic spin configuration in tetralayer MnBi$_2$Te$_4$ can be manipulated by pinning the spin orientations of the surface,  allowing the system to stabilize in
a new configuration without any net magnetization, but with a nonvanishing Chern number  \cite{PhysRevLett.134.116603}.

Complicating matters further, the application of magnetic fields may alter the Hall conductivity of anomalous quantum Hall materials. Reports have indicated the emergence of a higher Chern-number phase arising from the coexistence of the quantum Hall effect and the Chern insulator phase in thin films of MnBi$_2$Te$_4$ under strong magnetic fields. In this regime, the lowest-Landau level is filled and forms a quantum Hall edge state that contributes an additional $e^2/h$ to the Hall conductivity \cite{PhysRevLett.127.236402,10.1093/nsr/nwac296,cai_electric_2022, 10.1093/nsr/nwac296,PhysRevB.108.125424} 
\begin{equation}
    \sigma_{\mathrm{H}} = (C_{\mathcal{V}} + \nu) \frac{e^2}{h},
    \label{eq:qhe}
\end{equation}
where $e$ is the charge of an electron, $h$ is Planck's constant, $C_{\mathcal{V}}$ is the Chern number, and $\nu$ is the Landau filling factor. However, other studies indicate that the Chern number of antiferromagnetic MnBi$_2$Te$_4$ thin films in the absence of a magnetizing field is constrained to $|C_{\mathcal{V}}|\leq 1$ regardless of thickness \cite{PhysRevB.102.241406,PhysRevLett.122.107202,10.1093/nsr/nwaa089,doi:10.1021/acs.nanolett.2c02034}. 

The linear response of MnBi$_2$Te$_4$ thin films and its connection to topological and geometric properties has attracted significant interest. In particular, the conventional Kubo-Greenwood formula for the linear conductivity has been applied within a simplified coupled Dirac-cone model to up to ten SL layers \cite{PhysRevB.108.125424, PhysRevMaterials.5.064201}, with predictions for large Kerr and Faraday effects in response to a low-frequency electric field, for
films with an odd number of layers; these vanish for even-numbered layers. 
A more accurate approach using a Wannier-function-based tight-binding framework, though limited to three SL layers \cite{doi:10.1126/sciadv.ado1761}, revealed that the linear response is responsible for an enhanced magnetic circular dichroism within an infrared window, an effect not observed to date in any other material. An accurate treatment for a greater number of layers could reveal
responses with novel applications.

                In this work, we calculate the topological properties and linear response of MnBi$_2$Te$_4$ thin films within the independent particle approximation for the response, using recently derived expressions from a microscopic theory of polarization and magnetization
\cite{PhysRevB.111.075202,mahon2023polarization,kattan2025chern,PhysRevB.99.235140}. In this formalism, the linear response of a Chern insulator to a %n external 
finite-frequency electric field separates into a Kubo term associated with the dynamical (finite-frequency) response and a term describing the quantum anomalous Hall effect \cite{PhysRevB.111.075202}. Using
density functional theory (DFT) and ``single-shot" Wannier functions that preserve the symmetry \cite{m6c2-yd5j}, we numerically compute both contributions in thin films of one, four, five, and eleven SLs of MnBi$_2$Te$_4$, and identify the frequency regimes where the competing terms dominate. Furthermore, we examine topological signatures in these systems, such as Chern invariants and band inversions in the band structure, which typically arise from strong spin-orbit coupling (SOC) \cite{PhysRevB.98.235160,doi:10.1021/acs.jctc.5c00838}. 

\section{Computations}
First-principles density functional theory calculations were performed using the projector augmented-wave method \cite{PhysRevB.50.17953, PhysRevB.59.1758} and the Vienna Ab initio Simulation Package (VASP)~\cite{PhysRevB.47.558, PhysRevB.54.11169, KRESSE199615}. The fully relativistic Perdew-Burke-Ernzerhof (PBE) exchange-correlation functional was used~\cite{PhysRevLett.77.3865}, with a plane-wave cutoff energy of 600 eV. To treat the localized 3d-orbitals of Mn, the GGA$+U$ method was implemented with a $U$ parameter of 5.34 eV \cite{doi:10.1073/pnas.2207681119} using Dudarev's approach
\cite{PhysRevB.57.1505}.

\begin{figure}[t]
    \centering  \includegraphics[width=\linewidth]{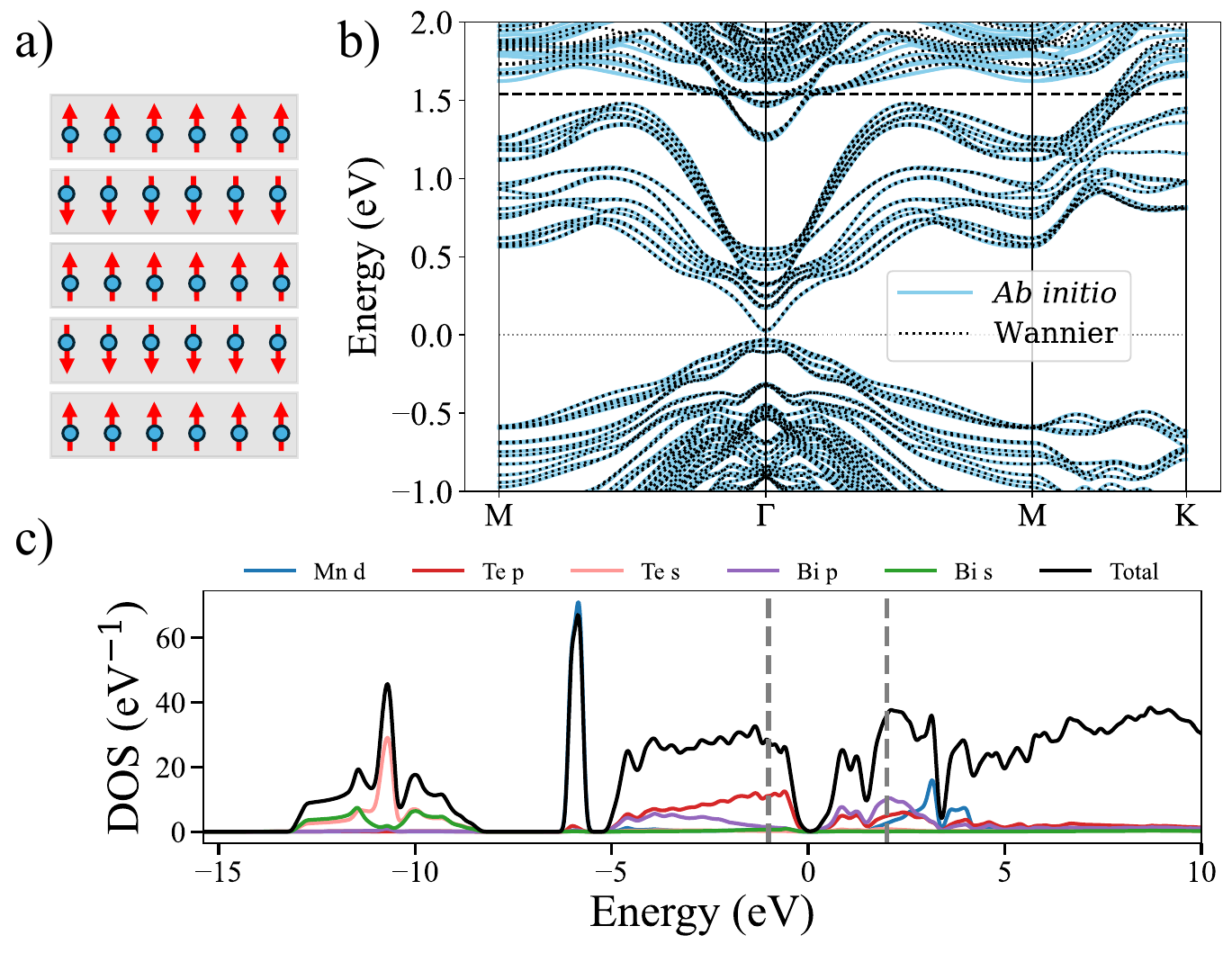}
    \caption{(a) Simplified lattice structure of 5 SL MnBi$_2$Te$_4$, where blue circles represent the Mn atom with spin magnetic moments. (b) The \textit{ab initio} and Wannier-interpolated band structure of 5 SL MnBi$_2$Te$_4$ along a high-symmetry path.  The Fermi energy is centered at 0 eV and the horizontal dashed line at $1.54$ eV denotes the frozen energy window maximum used in the disentanglement step in the Wannier construction. (c) Density of states for 5 SL MnBi$_2$Te$_4$; grey dashed vertical lines mark the energy window shown in (b).}
    \label{fig:structure}
\end{figure}

The bulk
MnBi$_2$Te$_4$ crystal structure belongs to the  $R\bar{3}m$ (no. 166) space group \cite{doi:10.1021/acs.chemmater.8b05017}. Fig.~\ref{fig:structure}(a) displays a simplified schematic of the layered structure composed of SLs that are held together by weak van der Waals forces, where Mn atoms carry spin magnetic moments with opposing directions as one moves from layer to layer. Films with an odd 
number of SLs preserve inversion symmetry, but break time-reversal symmetry, while those with an even number of SLs break both inversion and time-reversal symmetries, but preserve parity-time symmetry \cite{doi:10.1126/sciadv.ado1761}. The DFT band structure for 5 SLs is shown in Fig.~\ref{fig:structure}(b) alongside that obtained from Wannier interpolation. They are in good agreement within the frozen energy window, as discussed in more detail in the following section. Fig~\ref{fig:structure}(c) displays the density of states as a function of energy. At low energies, there are isolated bands (islands) composed predominantly of Bi and Te s-orbital character, followed by an island composed almost entirely of Mn d-orbitals. Near the Fermi energy, the valence bands are dominated by Te p-orbitals, while the conduction bands are dominated by Bi p-orbitals. The band gaps for $N=1, 4, 5$ and $11$ SLs were found to be 337, 35, 51, and 22 meV, respectively. Further computational details are provided in Appendix \ref{sec:WannierDetails}

\section{Wannier interpolation scheme}
Here we outline the Wannier interpolation scheme, which interpolates quantities calculated on a coarse reciprocal-space grid onto a fine grid, a step crucial for numerical convergence. We begin with the Bloch energy eigenfunctions for 2D systems,
\begin{equation}
    \psi_{n \boldsymbol{k}}(\boldsymbol{x})=\frac{1}{2\pi}e^{i\boldsymbol{k}\cdot \boldsymbol{x}}u_{n\boldsymbol{k}}(\boldsymbol{x}),
\end{equation}
which are solutions of the periodic Hamiltonian under Bloch's theorem \cite{bloch_uber_1929}, where $u_{n\boldsymbol{k}}(\boldsymbol{x})$ are cell-periodic two-component spinors, 
$u_{n\boldsymbol{k}}(\boldsymbol{x+R})=u_{n\boldsymbol{k}}(\boldsymbol{x})$ for all lattice vectors $\boldsymbol{R}$. One can construct the unitary transformation 
\begin{equation}
    u_{\alpha \boldsymbol{k}}(\boldsymbol{x}) = \sum_{n}U_{n\alpha}(\boldsymbol{k}) u_{n\boldsymbol{k}}(\boldsymbol{x}),
    \label{eq:smooth_Bloch}
\end{equation}
where the sum is over the bands that are to be Wannierized and the index $\alpha \in \mathbb{N}$ is generally distinct from the band index $n\in \mathbb{N}.$ The quantities $U_{n\alpha}(\boldsymbol{k})$ are the matrix components of a $\boldsymbol{k}$-dependent unitary operator $U(\boldsymbol{k}).$ From Eq.~(\ref{eq:smooth_Bloch}), one can construct Wannier functions
\begin{equation}
    W_{\alpha \boldsymbol{R}}(\boldsymbol{x}) = \sqrt{\Omega_{uc}} \int_{\text{BZ}^2} \frac{d\boldsymbol{k}}{(2\pi)^2} e^{i\boldsymbol{k}\cdot (\boldsymbol{x}-\boldsymbol{R})} u_{\alpha \boldsymbol{k}}(\boldsymbol{x}),
    \label{eq:Wannier}
\end{equation}
where $\Omega_{uc}$ is the volume of the unit cell. The gauge freedom in $U(\boldsymbol{k})$ can be chosen such that the resulting Wannier functions are maximally localized in real space. This corresponds to selecting $U(\boldsymbol{k})$ so that $u_{\alpha \boldsymbol{k}}(\boldsymbol{x})$ varies smoothly across the Brillouin zone (BZ) and the spread functional is minimized \cite{RevModPhys.84.1419}. From the outputs of Wannier90, two important quantities %are 
provided are $\braket{\boldsymbol{0} \alpha | \hat{H} | \boldsymbol{R} \beta}
$ and $\braket{\boldsymbol{0} \alpha | \hat{\boldsymbol{r}} | \boldsymbol{R}\beta},$ where $W_{\beta \boldsymbol{R}}(\boldsymbol{x})=\braket{\boldsymbol{x}|\boldsymbol{R}\beta }$. These matrix elements can then be used to compute the necessary quantities that we will need via Wannier interpolation \cite{PhysRevB.74.195118}, namely 
\begin{equation}
    \tilde{H}_{\alpha \beta} (\boldsymbol{k}) = \sum_{\boldsymbol{R}}e^{i\boldsymbol{k}\cdot \boldsymbol{R}} \braket{\boldsymbol{0} \alpha | \hat{H} | \boldsymbol{R} \beta},
    \label{eq:Htilde_ab}
\end{equation}
\begin{equation}
    \tilde{\xi}^\ell_{\alpha \beta} (\boldsymbol{k}) = \sum_{\boldsymbol{R}}e^{i\boldsymbol{k}\cdot \boldsymbol{R}} \braket{\boldsymbol{0} \alpha | \hat{r}_\ell | \boldsymbol{R}\beta},
    \label{eq:Xitilde_ab}
\end{equation}
where $\hat{H}$ is the crystal Hamiltonian. 
Diagonalizing $\tilde{H}(\boldsymbol{k})$ via $U(\boldsymbol{k}) \tilde{H} (\boldsymbol{k})U^\dagger(\boldsymbol{k})=H(\boldsymbol{k})$ yields matrix elements $H_{nm}(\boldsymbol{k})=E_{n\boldsymbol{k}} \delta_{nm},$ where the eigenvalues $E_{n\boldsymbol{k}}$ match the Bloch energies accurately within the frozen energy window, as displayed in Fig.~\ref{fig:structure}(b). Geometric quantities, such as the Berry connection in the Bloch basis, can also be accurately interpolated, and are related to their counterparts in the Wannier basis via
\begin{equation}
\sum_{\alpha\beta} U_{m\beta}(\boldsymbol{k}) \, \tilde{\xi}^i_{\beta\alpha}(\boldsymbol{k}) \, U^\dagger_{\alpha n}(\boldsymbol{k})
= \xi^i_{mn}(\boldsymbol{k}) + \mathcal{W}^i_{mn}(\boldsymbol{k}),
\label{eq:Berry_Bloch_Jasons}
\end{equation}
where we have introduced the Hermitian matrix elements
\begin{equation}
\mathcal{W}^i_{mn}(\boldsymbol{k}) \equiv 
i \sum_{\alpha} (\partial_i U_{m\alpha}(\boldsymbol{k})) U^\dagger_{\alpha n}(\boldsymbol{k}).
\label{eq:Ws}
\end{equation}
Eqs.~(\ref{eq:Berry_Bloch_Jasons}) and (\ref{eq:Ws}) follow the conventions used in our earlier work \cite{mahon2023polarization,kattan2025chern}. However, because the notation adopted in the Wannier90 community differs slightly \cite{PhysRevB.75.195121}, we include a footnote to clarify the relationship between both conventions, and show explicitly how to calculate the Berry connection in the Bloch basis from Eq.~(\ref{eq:Xitilde_ab}) \footnote{\label{fn:berry_connection}In the Wannier interpolation literature, the Berry connection in the Bloch basis is most commonly written as \begin{equation}
    \xi_{\ell, nm} (\boldsymbol{k})=U_{n\alpha}(\boldsymbol{k})\tilde{\xi}_{\ell, \alpha \beta}(\boldsymbol{k}) U^{\dagger}_{\beta m}(\boldsymbol{k}) + i U_{n\alpha}(\boldsymbol{k}) \partial_{\ell} U^{\dagger}_{\alpha m}(\boldsymbol{k}).
    \label{eq:BlochBerry}
\end{equation}
The second term in Eq.~(\ref{eq:BlochBerry}) can be evaluated using $\boldsymbol{k}\cdot \boldsymbol{p}$ perturbation theory and is related to the anti-Hermitian matrix 
\begin{equation}
(U(\boldsymbol{k}) \partial_{\ell} U^{\dagger}(\boldsymbol{k}))_{nm}
= 
\begin{cases}
\dfrac{\bar{H}_{\ell, nm}(\boldsymbol{k})}{E_{m \boldsymbol{k}} - E_{n \boldsymbol{k}}}, & \text{if } n \neq m, \\[1em]
0, & \text{if } n = m
\label{eq:Ds}
\end{cases}
\end{equation}
where we have defined $\bar{H}_{\ell}(\boldsymbol{k})=U (\boldsymbol{k})\tilde{H}_{\ell}(\boldsymbol{k})U^\dagger(\boldsymbol{k})$, etc. Note that Eq.~(\ref{eq:Ds}) diverges in the case of band crossings. However, here we only consider matrix elements between unoccupied and occupied states of an insulator.}\setcounter{berryfoot}{\value{footnote}}.
 The advantages of Wannier interpolation become clear in Eqs.~(\ref{eq:Htilde_ab}) and (\ref{eq:Xitilde_ab}), where for Wannier functions that are very well localized the matrix elements $\braket{\boldsymbol{0} \alpha | \hat{H} | \boldsymbol{R} \beta}
$ and $\braket{\boldsymbol{0} \alpha | \hat{\boldsymbol{r}} | \boldsymbol{R}\beta}$ decay rapidly with increasing $|\boldsymbol{R}|$ \cite{RevModPhys.84.1419}. Consequently, only a small number of lattice vectors need be included in the summations to interpolate both the band structure and geometric quantities accurately. 
\begin{figure*}[t]
    \centering  \includegraphics[width=0.9\textwidth]{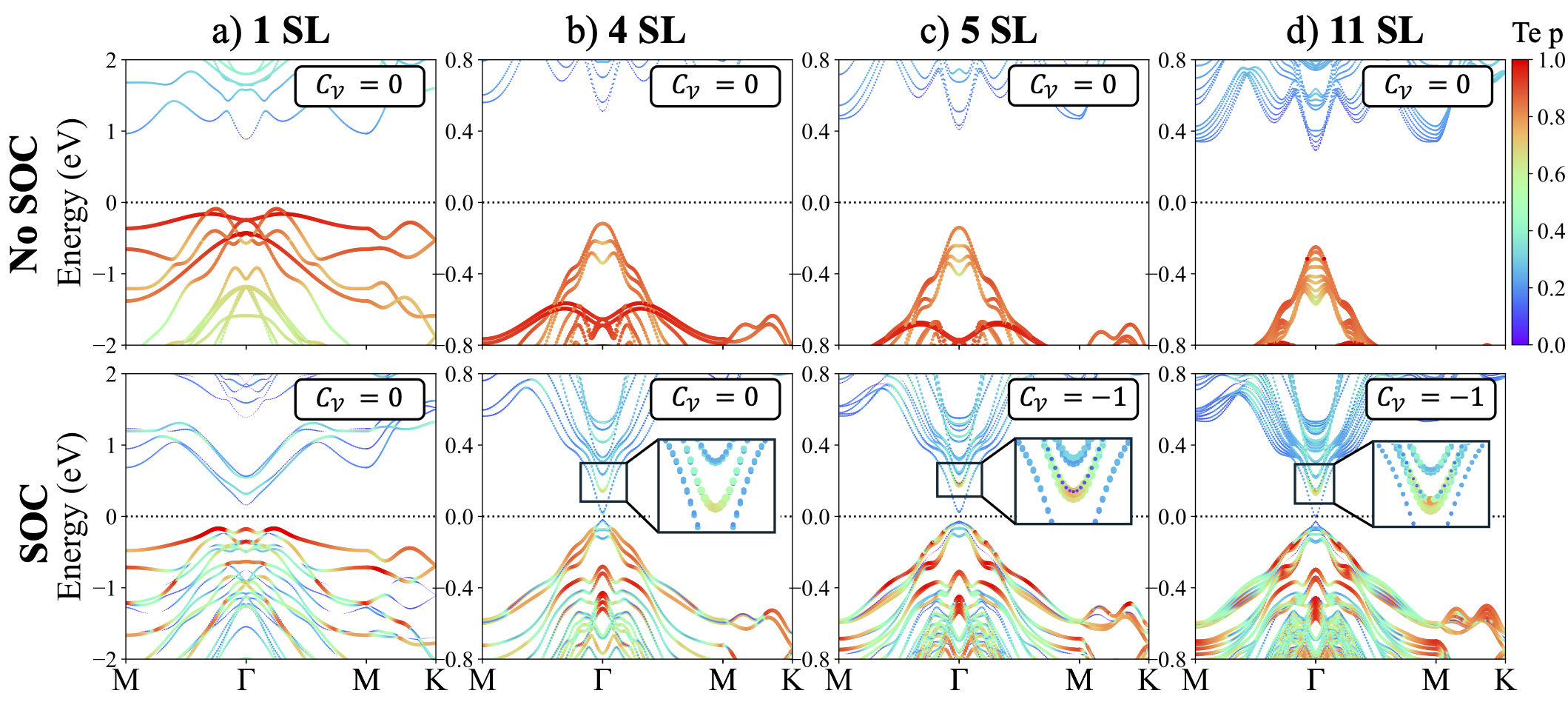}
    \caption{Orbital resolved band structure identifying the contributions of Te p states in MnBi$_2$Te$_4$ thin films of (a) 1 SL, (b) 4 SL, (c) 5 SL, and (d) 11 SL. Band colors indicate Te p-orbital weight (scale shown in the upper-right panel of (d)). No topological band inversion occurs in 1 SL upon the inclusion of SOC, whereas it appears in 4, 5, and 11 SL systems. The corresponding Chern numbers are shown in the upper-right corner of each panel. Upper panels are calculated without SOC, while lower panels include SOC.}
    \label{fig:bandinversion}
\end{figure*}

\section{Topological signatures}
The Chern number characterizes the twisting of the Bloch states as they are parallel transported along curves in the Brillouin zone \cite{mahon2023polarization}.
The most familiar expression for the Chern number involves integrating the Berry curvature of the occupied bands over the BZ \cite{vanderbilt2018berry}
\begin{equation}
    C_{\mathcal{V}} = \frac{1}{2\pi} \sum_{n} f_{n}\int_{\text{BZ}^2} d\boldsymbol{k} \, \varepsilon^{ab} \partial_{a} \xi^b_{nn} (\boldsymbol{k}),
\end{equation}
where the Levi-Civita symbol is defined by $\varepsilon^{xy}=-\varepsilon^{yx}=1$ and $\varepsilon^{xx}=\varepsilon^{yy}=0$. The filling factor $f_n$ equals 1 if the band indexed by $n$ is occupied and 0 otherwise. In the conventional approach, the Berry curvature is evaluated using expressions that involve inverse powers of band-energy differences \cite{PhysRevB.74.195118}. Near band crossings, it is necessary to define an artificial energy tolerance when performing numerical calculations, below which contributions from small energy differences are neglected. To circumvent this, it is possible to express the Chern number in a form that avoids these matrix elements. Recently
\cite{kattan2025chern}, it was shown that the Chern number is given by
\begin{equation}
    C_{\mathcal{V}} = \frac{h^2}{4\pi i} \sum_{mn} f_{nm}\int_{\text{BZ}^2} \frac{d\boldsymbol{k}}{(2\pi)^2} \frac{\varepsilon^{ab} {v}^a_{nm}(\boldsymbol{k}){v}^b_{mn}(\boldsymbol{k})}{(E_{n \boldsymbol{k}}-E_{m \boldsymbol{k}})(E_{m \boldsymbol{k}}-E_{n \boldsymbol{k}})},
    \label{eq:ChernNumber}
\end{equation}
 involving the matrix elements 
 \begin{equation}
    {\boldsymbol{v}}_{nm}(\boldsymbol{k}) = \frac{1}{\Omega_{\text{uc}}} \int_{\Omega} d\boldsymbol{x} \, u^\dagger_{n\boldsymbol{k}}(\boldsymbol{x}) \boldsymbol{v}(\boldsymbol{x}) u_{m\boldsymbol{k}}(\boldsymbol{x}), 
    \label{eq:velocity}
\end{equation}
 of the velocity operator ${\boldsymbol{v}}(\boldsymbol{x})$. The velocity matrix elements are related to the Bloch energies and Berry connections by  
\cite{PhysRevB.98.214402}
\begin{equation}
    v^a_{nm}(\boldsymbol{k})= \frac{\delta_{nm}}{\hbar} \frac{\partial E_{m\boldsymbol{k}}}{\partial k^a}+\frac{i}{\hbar} (E_{n\boldsymbol{k}}-E_{m\boldsymbol{k}})\xi^a_{nm}(\boldsymbol{k}).
\end{equation}

 Importantly, Eq.~(\ref{eq:ChernNumber}) is well-behaved when $n=m$ and around degeneracies, since the prefactor $f_{nm}=f_n-f_m$ vanishes, thereby canceling contributions from the poles in the denominator when $E_{n\boldsymbol{k}}=E_{m\boldsymbol{k}}$. Nonetheless, its evaluation requires including all bands; any truncation yields an approximate expression. It has also been shown that in the presence of time-reversal symmetry the Chern number vanishes, but it need not vanish in the presence of inversion symmetry alone \cite{kattan2025chern}. We will use these expressions to evaluate $C_\mathcal{V}$ (\ref{eq:ChernNumber}) for MnBi$_2$Te$_4$ thin films with $N = 1, 4, 5,$ and $11$ SLs to determine their topological character and reveal 
 band inversions driven by SOC, which are usually indicative of topological phase transitions. In all cases, we achieve convergence using a $300\times 300 \times 1$ $\boldsymbol{k}$-point interpolation mesh and adaptive mesh refinement consisting of 60 iterations, with the Hilbert space spanned by bands of Bi p-orbitals and Te p-orbitals around the Fermi level.

For a single-SL MnBi$_2$Te$_4$ thin film, we find that the Chern number is zero to good numerical accuracy (see Table \ref{tab:chern_numbers} in  Appendix \ref{sec:WannierDetails}), classifying it as a topologically trivial insulator, in agreement with previous reports \cite{doi:10.1126/sciadv.aaw5685}. As shown in the upper panel of Fig.~\ref{fig:bandinversion}(a), in the absence of SOC the valence bands are predominantly of Te p-orbital character, while the conduction bands are mainly of Bi p-orbital character. The band gap decreases significantly upon the inclusion of SOC, but the orbital characters remain unchanged, indicating the absence of any band inversions. Broken time-reversal symmetry in this system alone does not guarantee a nonzero Chern number, as a topologically trivial ferromagnet illustrates: The breaking of time-reversal symmetry is a necessary but not sufficient condition for a topologically nontrivial phase. 

The Chern number also vanishes in the case of 4 SLs, but the inclusion of SOC results in an apparent band inversion around $\Gamma$ as depicted in Fig.~\ref{fig:bandinversion}(b), where the Te p-orbital character from the valence band maximum has been transferred to bands slightly above the conduction band minimum. The dominant orbital character of the valence band maximum is now of Bi p-orbital character, inherited from the conduction band minimum. This swapping of orbital character between the valence and conduction bands, with a vanishing Chern number, indicates a 
phase distinct from the 1 SL system, and is found in Bi$_2$Se$_3$, Bi$_2$Te$_3$, and Sb$_2$Te$_3$, which are known $\mathbb{Z}_2$ topological insulators \cite{zhang_topological_2009}.

The Chern number for 5 SL was calculated to be $C_{\mathcal{V}}=-1$, indicating a Chern insulator phase \footnote{Others have reported a Chern number of $+1$ in 5 SL MnBi$_2$Te$_4$  \cite{PhysRevResearch.7.023024,PhysRevB.108.125424}. This difference in sign arises from their choice of convention in the Kubo-Greenwood formula, rather than a ground-state calculation as in Eq.~(\ref{eq:ChernNumber})}. The well-known topological obstruction that prevents the construction of exponentially localized Wannier functions can be overcome by including unoccupied states so that the net Chern number of the bands used in calculating the Wannier functions is zero \cite{PhysRevLett.121.126402}. The Chern insulator phase has been %is 
attributed to the combination of the exchange splitting from the ferromagnetic SLs due to the Mn atoms and the band inversion around $\Gamma$ from the strong SOC \cite{doi:10.1126/sciadv.aaw5685}, as shown in the lower panel of Fig.~\ref{fig:bandinversion}(c). We confirm that, in the absence of SOC, the Chern number of 5 SL drops to zero with high numerical accuracy, as shown in Table \ref{tab:chern_numbers} of Appendix \ref{sec:WannierDetails}. 

Higher Chern-number phases with ``Chern number" of $\pm 2$ have been reported in 9 and 10 SLs MnBi$_2$Te$_4$ thin films, but under a perpendicular applied magnetic field where the magnetic moments between layers align ferromagnetically \cite{10.1093/nsr/nwaa089}. For the 11 SL film, we observe a SOC-induced band inversion as shown in Fig.~\ref{fig:bandinversion}(d), and we obtain a Chern number $C_{\mathcal{V}}=-1$, suggesting 
that the observation of phases with higher Chern numbers could be a consequence of a combination of the quantum anomalous Hall effect and the quantum Hall effect, as in Eq.~(\ref{eq:qhe}). We emphasize that our calculations are performed in the zero-field A-type antiferromagnetic configuration; therefore, our result of $C_\mathcal{V}=-1$ in the 11 SL system does not contradict the phases with higher Chern numbers reported in the 9 and 10 SL films that arise from a field-driven ferromagnetic configuration. Our findings are consistent with a $\boldsymbol{k}\cdot \boldsymbol{p}$ model for MnBi$_2$Te$_4$ thin films, which predicts $C_{\mathcal{V}}=-1$ for films with an odd number of SLs greater than or equal to three, and vanishing Chern number for even layered films \cite{PhysRevB.102.241406}.

\section{Linear response}
Up to this point, we have focused on
the properties of the ground state. We now
turn to the long-wavelength response of MnBi$_2$Te$_4$ SLs to light. We begin by briefly outlining the formalism in which the linear response of Chern insulators was derived. We define microscopic polarization and magnetization fields $(\boldsymbol{p},\boldsymbol{m})$, as well as free charge and current densities $({\rho_{F},\boldsymbol{j}_F})$ that satisfy the relation \cite{PhysRevB.99.235140}
\begin{align}
\rho(\boldsymbol{x},t) &= -\boldsymbol{\nabla} \cdot 
\boldsymbol{p}(\boldsymbol{x},t) + \rho_{F}(\boldsymbol{x},t), \nonumber \\
\boldsymbol{j}(\boldsymbol{x},t) &= \frac{\partial \boldsymbol{p}(\boldsymbol{x},t)}{\partial t}
+ c \boldsymbol{\nabla} \times \boldsymbol{m}(\boldsymbol{x},t)
+ \boldsymbol{j}_F(\boldsymbol{x},t).
\end{align}
These fields are defined through the Wannier functions in Eq.~(\ref{eq:Wannier}), and the corresponding macroscopic fields can be obtained through spatial averaging \cite{PhysRevB.111.075202}. Through a perturbative expansion of these microscopic fields in powers of the wavevector of the electromagnetic field, one can calculate the long wavelength response of the corresponding macroscopic quantities. At this order the magnetization does not contribute, and the linearly induced macroscopic current density is given by
\begin{align}
    \boldsymbol{J}^{(1)}(\boldsymbol{x},t) = \frac{\partial \boldsymbol{P}^{(1)}(\boldsymbol{x},t)}{\partial t} + \boldsymbol{J}_F^{(1)}(\boldsymbol{x},t).
    \label{eq:J1}
\end{align}
The conductivity relates this macroscopic current density in $d=3$ dimensions ($i, \ell \in \{x,y,z\})$ to the total electric field $E^{i} (\omega)$ via \cite{kattan2025chern} 
\begin{equation}
    {J}^{(1)i}(\omega) = \sigma^{i \ell}_{3\text{D}}(\omega) E^{\ell} (\omega),
\end{equation}
or in $d=2$ dimensions $(i, \ell \in \{x,y \})$
\begin{equation}
    K^{(1)i}(\omega) = \sigma^{i \ell}_{2\text{D}} (\omega) E^{\ell}_{\parallel}(\omega),
\end{equation}
where $\boldsymbol{E}_{\parallel}(\omega)$ is the spatially uniform electric field lying in the plane of the sheet, and the conductivity tensor is the sum of two terms 
\begin{equation}
    \sigma^{i\ell}_{2\text{D}} (\omega) = \sigma_{\mathrm{K}}^{i\ell}(\omega) - \frac{e^2}{2\pi \hbar} \varepsilon^{i\ell} C_\mathcal{V}, 
    \label{eq:conductivity}
\end{equation}
where the first term, known as the Kubo term, is associated with the finite-frequency response and also arises for
topologically trivial insulators \cite{PhysRevResearch.2.043110}; it is 
given by 
\begin{widetext}
\begin{equation}
\sigma_{\mathrm{K}}^{i\ell}(\omega)
= -i \omega e^2 \hbar^2
\sum_{mn} f_{nm} 
\int_{\mathrm{BZ}^2} \frac{d\boldsymbol{k}}{(2\pi)^2} \frac{1}{(E_{m\boldsymbol{k}} - E_{n\boldsymbol{k}})^2}
\frac{{v}^i_{nm}(\boldsymbol{k})\, {v}^\ell_{mn}(\boldsymbol{k})}
{E_{m\boldsymbol{k}} - E_{n\boldsymbol{k}} - \hbar\omega - i\eta},
\label{eq:Kubo}
\end{equation}
\end{widetext}

\begin{figure*}[t]
    \centering  \includegraphics[width=0.8\textwidth]{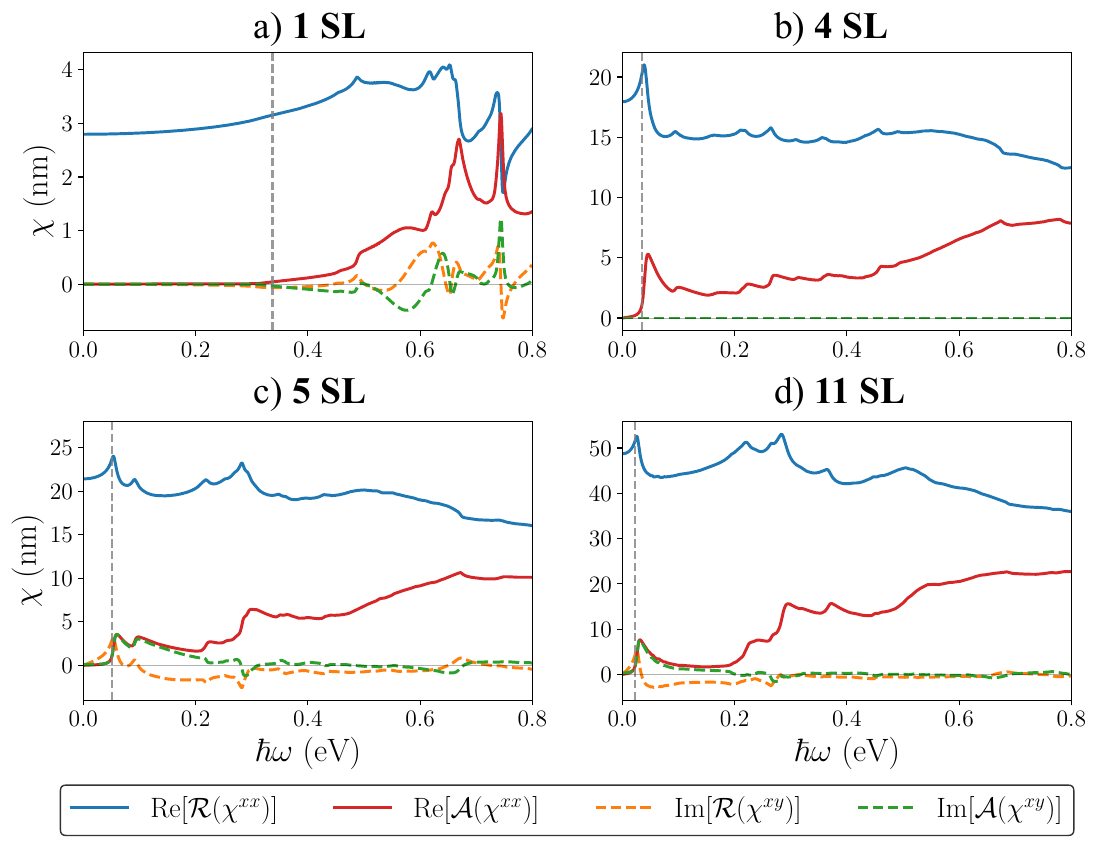}
    \caption{Reactive $(\mathcal{R})$ and absorptive $(\mathcal{A})$ components of the two-dimensional Kubo susceptibility for thin films of (a) 1 SL, (b) 4 SL, (c) 5 SL, and (d) 11 SL MnBi$_2$Te$_4$. In the 4 SL system, the combination of parity-time symmetry and $C_{3z}$ rotational symmetry ensures that $\mathcal{R}(\chi^{xy})$ and $\mathcal{A}(\chi^{xy})$ vanish at all frequencies. The grey dashed vertical line indicates the band gap energy. The
    real and imaginary parts of the tensor components that are not shown vanish at all frequencies. See Appendix \ref{sec:Ku} for further discussion.}
    \label{fig:chi}
\end{figure*}

\noindent where $\eta>0$ is introduced as a broadening parameter; we set it to 3 meV in our calculations. This Kubo conductivity originates from the first-order polarization field as in Eq.~(\ref{eq:J1}) and contributes to both the longitudinal and transverse components of the total current density (\ref{eq:J1}). The second term in Eq.~(\ref{eq:conductivity}) involves the Chern number $C_\mathcal{V}$, and encodes the Hall conductivity. This topologically quantized Hall conductivity arises from the transverse component of the free current density on the right-hand-side of Eq.~(\ref{eq:J1}). We note that a quantized Hall conductivity observed in transport does not necessarily imply an insulating bulk, as it may coexist with a gapless metallic bulk and a finite longitudinal conductivity \cite{8vs2-jvc4}. By decomposing the total conductivity into Kubo and Hall contributions, one can easily determine whether the non-topological or topological term is dominant across different frequency ranges. In the static limit, %where 
$\omega\rightarrow 0$, the Kubo term vanishes, leaving only the Hall contribution, which gives rise to the quantum anomalous Hall effect. Both the Kubo term and the total conductivity tensors satisfy $\sigma^{xx}(\omega)=\sigma^{yy}(\omega)$ and $\sigma^{xy}(\omega)=-\sigma^{yx}(\omega)$, as enforced by the $C_{3z}$ in-plane rotational symmetry present in MnBi$_2$Te$_4$ thin films for all number of SLs \cite{doi:10.1126/sciadv.ado1761}. The Kubo conductivities for 1, 4, 5, and 11 SLs are shown in the upper panels of Figs.~\ref{fig:KuboCond}(a-d) and discussed in detail in Appendix \ref{sec:Ku}.

We can also define a Kubo susceptibility that is related to the Kubo conductivity by $\chi^{i \ell}(\omega)= i\sigma_{\mathrm{K}}^{i \ell} (\omega)/{\omega}$.  From the Sokhotski–Plemelj identity 
\begin{equation}
    \lim_{\eta \rightarrow 0^+} \frac{1}{\omega' -\omega \pm i \eta} = \text{p.v.} \left(\frac{1}{\omega' - \omega}\right) \mp i\pi \delta(\omega' - \omega),
\end{equation}
 the Kubo susceptibility can be separated into a reactive (Hermitian) piece and an absorptive (anti-Hermitian) piece \cite{lifshitz1995physical}. The reactive piece is given by
\begin{widetext}
\begin{equation}
    \mathcal{R}(\chi^{i \ell}(\omega)) = e^2 \hbar^2 \sum_{mn} f_{nm} \fint_{\mathrm{BZ}^2} \frac{d\boldsymbol{k}}{(2\pi)^2} \frac{1}{(E_{m\boldsymbol{k}} - E_{n\boldsymbol{k}})^2}
\frac{{v}^i_{nm}(\boldsymbol{k})\, {v}^\ell_{mn}(\boldsymbol{k})}
{E_{m\boldsymbol{k}} - E_{n\boldsymbol{k}} - \hbar\omega},
\end{equation}
where the bar on the integral indicates
the Cauchy principal part, while the absorptive piece is given by 
\begin{equation}
    \mathcal{A}(\chi^{i \ell}(\omega)) = \pi e^2 \hbar^2 \sum_{mn} f_{nm} \int_{\mathrm{BZ}^2} \frac{d\boldsymbol{k}}{(2\pi)^2}
\frac{{v}^i_{nm}(\boldsymbol{k})\, {v}^\ell_{mn}(\boldsymbol{k})}
{(E_{m\boldsymbol{k}} - E_{n\boldsymbol{k}})^2}
\delta(E_{m\boldsymbol{k}} - E_{n\boldsymbol{k}} - \hbar\omega).
\label{eq:Chi_absorptive}
\end{equation}
\end{widetext}
The reactive and absorptive components of the susceptibility are related to the full tensor by 
\begin{equation}
\begin{aligned}
\mathcal{R}\!\left(\chi^{i\ell}(\omega)\right)
&= \frac{1}{2}\Big(\chi^{i\ell}(\omega)+(\chi^{\ell i}(\omega))^*\Big), \\
\mathcal{A}\!\left(\chi^{i\ell}(\omega)\right)
&= \frac{1}{2i}\Big(\chi^{i\ell}(\omega)-(\chi^{\ell i}(\omega))^*\Big),
\end{aligned}
\label{eq:HermitianAnti}
\end{equation}
and satisfy the usual Kramers-Kronig relations \cite{lucarini2005kramers} even if time-reversal symmetry is absent, 
\begin{figure}[t]
    \centering  \includegraphics[width=\linewidth]{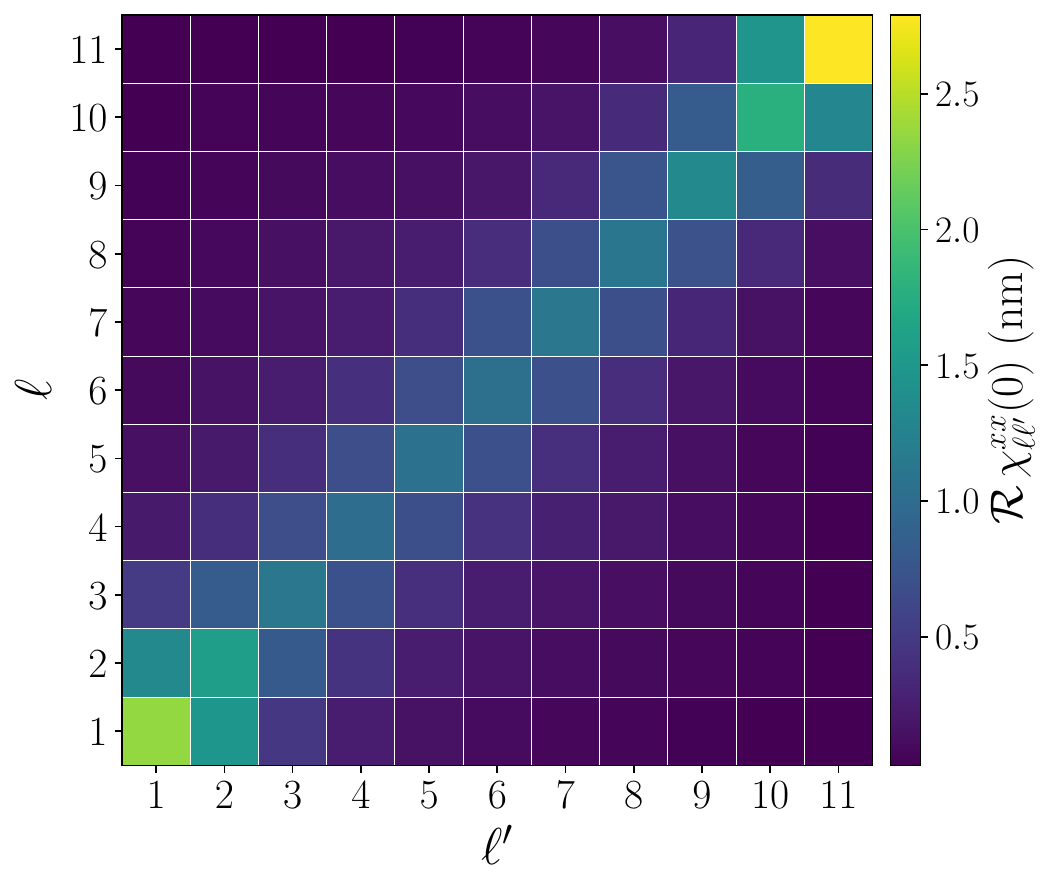}
    \caption{Layer-resolved decomposition of the static limit of the reactive susceptibility component $\mathcal{R}(\chi^{x x}_{\ell\ell'}(0))$ for 11 SL MnBi$_2$Te$_4$.}
    \label{fig:layer_resolved_chi}
\end{figure}
\begin{equation}
\begin{aligned}
    \mathcal{R}(\chi^{i \ell}(\omega)) &= \frac{1}{\pi} \fint_{-\infty}^\infty d\omega' 
    \frac{\mathcal{A}(\chi^{i \ell}(\omega'))}{\omega' - \omega}, \\
    \mathcal{A}(\chi^{i \ell}(\omega)) &= -\frac{1}{\pi} \fint_{-\infty}^\infty d\omega' 
    \frac{\mathcal{R}(\chi^{i \ell}(\omega'))}{\omega' - \omega}.
\end{aligned}
\end{equation}
The absorptive part of the Kubo susceptibility, given by Eq.~(\ref{eq:Chi_absorptive}),  
arise only for frequencies where $\hbar \omega$ exceeds the band gap.  The Hermiticity of the velocity matrix elements in Eq.~(\ref{eq:velocity}) ensures that the reactive and absorptive components of $\chi^{xx}(\omega)$ are purely real. Furthermore, again as a result of the $C_{3z}$ rotational symmetry, the Kubo susceptibility also satisfies $\chi^{xx}(\omega)=\chi^{yy}(\omega)$ and $\chi^{xy}(\omega)=-\chi^{yx}(\omega)$; using Eq.~(\ref{eq:HermitianAnti}), we see that the reactive and absorptive components of $\chi^{xy}(\omega)$ are purely imaginary. Figs.~\ref{fig:chi}(a-d) presents the reactive and absorptive components of the two-dimensional Kubo susceptibility for 1, 4, 5 and 11 SL MnBi$_2$Te$_4$, expressed in
the Gaussian system \cite{jackson1998appendix} of electromagnetic quantities and units \footnote{Conversion to the SI system of electromagnetic quantities and units is
achieved via the relation $\chi^{SI}= 4\pi \times10^{-2} \chi^{Gaussian}$}.

For all of the septuple-layer systems, %In all systems, 
the absorptive components vanish below the band gap (marked by the gray-dashed vertical lines), except for small contributions near the gap due to the broadening of the delta function. In contrast, $\mathcal{R}(\chi^{xx})$ remains finite at zero frequency and grows with thickness, showing a peak at the band gap in the 4, 5, and 11 SL systems. In the static limit, the reactive susceptibility component can be written as
\begin{equation}
    \mathcal{R}(\chi^{xx}(0)) = e^2  \sum_{mn} f_{nm} \int_{\mathrm{BZ}^2} \frac{d\boldsymbol{k}}{(2\pi)^2} 
\frac{g^{xx}_{nm}(\boldsymbol{k})}
{E_{m\boldsymbol{k}} - E_{n\boldsymbol{k}}},
\label{eq:reactive_quantum_metric}
\end{equation}
where $g^{xx}_{nm}(\boldsymbol{k})=\xi^{x}_{nm}(\boldsymbol{k}) \xi^x_{mn}(\boldsymbol{k})$ are the matrix elements of the non-Abelian quantum metric associated with virtual interband transitions. The form of Eq.~(\ref{eq:reactive_quantum_metric}) resembles that of the ``intrinsic capacitance" that governs the low-frequency response of the conductivity, as recently discussed in Komissarov \textit{et al.} \cite{komissarov_quantum_2024}. We find that the septuple layer dependence of $ \mathcal{R}(\chi^{x x}(0))$ is well parametrized by a linear function of the thickness of the layers as shown in Fig.~\ref{fig:MagCircDich}(c). To gain insight into the spatial distribution of the static reactive susceptibility component in $N_L$-SL MnBi$_2$Te$_4$, we introduce a layer-resolved decomposition (see Appendix \ref{sec:layer_resolved} for details)
\begin{equation}
    \label{eq:Layer_reactive_chi}
    \mathcal{R}(\chi^{x x}(0)) =   \sum_{\ell,\ell'=1}^{N_{L}} \mathcal{R}(\chi^{x x}_{\ell\ell'}(0)),
\end{equation}
where $\ell, \ell' \in \{1,\ldots,N_L\}$ denote the SL indices. Fig.~\ref{fig:layer_resolved_chi} shows $\mathcal{R}(\chi^{x x}_{\ell\ell'}(0))$ for the 11 SL system. The intralayer contributions are largest near the two surfaces (outer layers) and decrease toward the interior layers, indicating that the reactive response is non-uniformly distributed throughout the film. Notably, the off-diagonal weight is significant: the total interlayer contribution is about twice the total intralayer contribution. Overall, the contributions involving only the bulk layers ($\ell,\ell'\in \{2,...10\}$) account for roughly 70\% of the total response, while the remaining 30\% involves at least one surface layer.

Returning to the frequency-dependent response, in the 5 and 11 SL, $\mathcal{R}(\chi^{xy})$ shows a similar peak at the energy gap. Furthermore, in the 1, 5 and 11 SL systems, where time-reversal symmetry is broken, the off-diagonal components of susceptibility take on finite values. In contrast, in the 4 SL system parity-time symmetry enforces $\chi^{xy}(\omega)=\chi^{yx}(\omega)$, which together with $C_{3z}$ rotational symmetry implies that $\chi^{yx}(\omega)=0$ at all frequencies, consistent with the behavior reported for 2 SL MnBi$_2$Te$_4$ \cite{doi:10.1126/sciadv.ado1761}. 

The absorptive components play very important roles in the optical response: $\mathcal{A}^{xx}(\omega)$ is related to the absorption of linearly polarized light and
$\mathcal{A}^{xy}(\omega)$ is responsible for magnetic circular dichroism. For the 1 SL system, the slow increase in $\mathcal{A}(\chi^{xx})$ with frequency has been attributed to the absence of a topological band inversion \cite{doi:10.1126/sciadv.ado1761}. Unlike 1 SL, in the 4, 5 and 11 SL systems we observe a pronounced peak in $\mathcal{A}(\chi^{xx})$ as $\hbar \omega$ increases slightly above their respective energy gaps, arising from band inversion and enhanced absorption. 

As shown in Figs.~\ref{fig:chi}(c) and \ref{fig:chi}(d), the Kubo susceptibility of 5 and 11 SL satisfies Re$[\mathcal{A}(\chi^{xx})]\approx$ Im$[\mathcal{A}(\chi^{xy})]$ within the energy range $51 \leq \hbar\omega \leq  150$ meV, similar to the energy range $65 \leq \hbar \omega \leq 150$ meV reported in 3 SL MnBi$_2$Te$_4$ \cite{doi:10.1126/sciadv.ado1761}. Furthermore, within a simple two-band gapped Dirac cone model considered by B. Ghosh \textit{et al}. \cite{doi:10.1126/sciadv.ado1761}, the equality Re$[\mathcal{A}(\chi^{xx})] = |$Im$[\mathcal{A}(\chi^{xy})]|$ holds exactly at the band edge. Unlike for the 3 SL MnBi$_2$Te$_4$ system, experiments have confirmed that the 5 SL system hosts a Chern insulator phase \cite{doi:10.1126/science.aax8156}. This condition Re$[\mathcal{A}(\chi^{xx})]\approx$ Im$[\mathcal{A}(\chi^{xy})]$ produces a ``nearly perfect" magnetic circular dichroism within these narrow infrared energy windows, where only one circular polarization is absorbed \cite{doi:10.1126/sciadv.ado1761}. The 5 and 11 SL Kubo conductivities show the same behavior, where  Re$[\sigma_\mathrm{K}^{xx}]\approx$ Im$[\sigma_\mathrm{K}^{xy}]$, as in Figs.~\ref {fig:MagCircDich}(a) and \ref{fig:MagCircDich}(b).

\section{Summary}
In summary, we calculated the two-dimensional Chern numbers using Eq.~(\ref{eq:ChernNumber}) for 1, 4, and 5 SL MnBi$_2$Te$_4$ thin films to be $0$, $0$, and $-1$, respectively, in agreement with previous reports. We identified topological spin-orbit-coupling-driven band inversions in all systems except the 1 SL film. We found that the Chern number for the 11 SL system, which is a property of the occupied valence bands in the system's ground state, is $C_{\mathcal{V}}=-1$. This indicates that the reported ``higher Chern-number phases"  arise from additional effects such as the filling of Landau levels due to a magnetizing field or quantum confinement \cite{10.1093/nsr/nwac296}. For example, when a strong magnetizing field is applied, this can lead to the filling of one or more Landau levels ($\nu>0$) associated with the quantum Hall effect, in addition to the usual quantum anomalous Hall response associated with the Chern insulator phase  (see Eq.~(\ref{eq:qhe})) \cite{PhysRevLett.127.236402}.

Using recently derived expressions for the linear response of a Chern insulator to a frequency-dependent electric field, we calculated the optical conductivity and susceptibility for 1, 4, 5, and 11 
SL systems. These calculations provide quantitative predictions in the terahertz to infrared frequency range that are directly accessible via optical spectroscopy \cite{tomarchio2022electrodynamics}, and highlight the
thickness dependence of the response. In particular, the static reactive susceptibility shows a non-uniform spatial distribution: the largest intralayer contributions arise from the surface layers, the total interlayer contributions account for nearly twice the total intralayer contributions, and approximately 70\% of the total response is attributed to the bulk layers.  We also identified the effects of symmetries and topology in shaping their responses. In the 5 and 11 SL responses, we observe a similar infrared energy window with nearly perfect magnetic circular dichroism \cite{doi:10.1126/sciadv.ado1761}. In future work, we will use these band structures and response tensors to study optical properties, including Faraday and Kerr rotations, and the transmission and reflection of light across MnBi$_2$Te$_4$ thin films. Furthermore, we will extend our microscopic polarization and magnetization formalism to study the response in nonlinear regimes and the spin conductivity in $\mathbb{Z}_2$ topological insulators.

%%%%%%%%%%%%%%%%%%%%%%%%%%%%%%%%%%%%%%%%%%%%%%%%%%%%%%%%%%%%%%%%%%%%%%%%%%%%%%%%%%%%%%%%%%

%%%%%%%%%%%%%%%%%%%%%%%%%%%%%%%%%%%%% Acknowledgments %%%%%%%%%%%%%%%%%%%%%%%%%%%%%%%%%%%%

\section{Acknowledgments}
This work was supported by the Natural Sciences and Engineering Research Council of Canada (NSERC). M.A. acknowledges helpful conversations with V. Cr\'{e}pel and L. David Le Nir. Computations were performed on the Trillium cluster hosted by the Digital Research Alliance of Canada. MA is supported by the Ontario Graduate Scholarship (OGS). JS and JI-A acknowledge funding from the European Union’s Horizon 2020 research and innovation programme under the European Research Council (ERC) Grant Agreement No. 946629 StG PhotoNow.

%%%%%%%%%%%%%%%%%%%%%%%%%%%%%%%%%%%%%%%%%%%%%%%%%%%%%%%%%%%%%%%%%%%%%%%%%%%%%%%%%%%%%%%%%%

%%%%%%%%%%%%%%%%%%%%%%%%%%%%%%%%%%%%%%%% Contribution %%%%%%%%%%%%%%%%%%%%%%%%%%%%%%%%%%%%

%\section{Author Contributions}

%%%%%%%%%%%%%%%%%%%%%%%%%%%%%%%%%%%%%%%%%%%%%%%%%%%%%%%%%%%%%%%%%%%%%%%%%%%%%%%%%%%%%%%%%%

\section*{Data availability}

The data and code needed to reproduce the figures and findings of this work are publicly available on \href{https://github.com/matthew7879/MnBi2Te4-linear-response.git}{GitHub}.

\begin{figure}[b]
    \centering  \includegraphics[width=\linewidth]{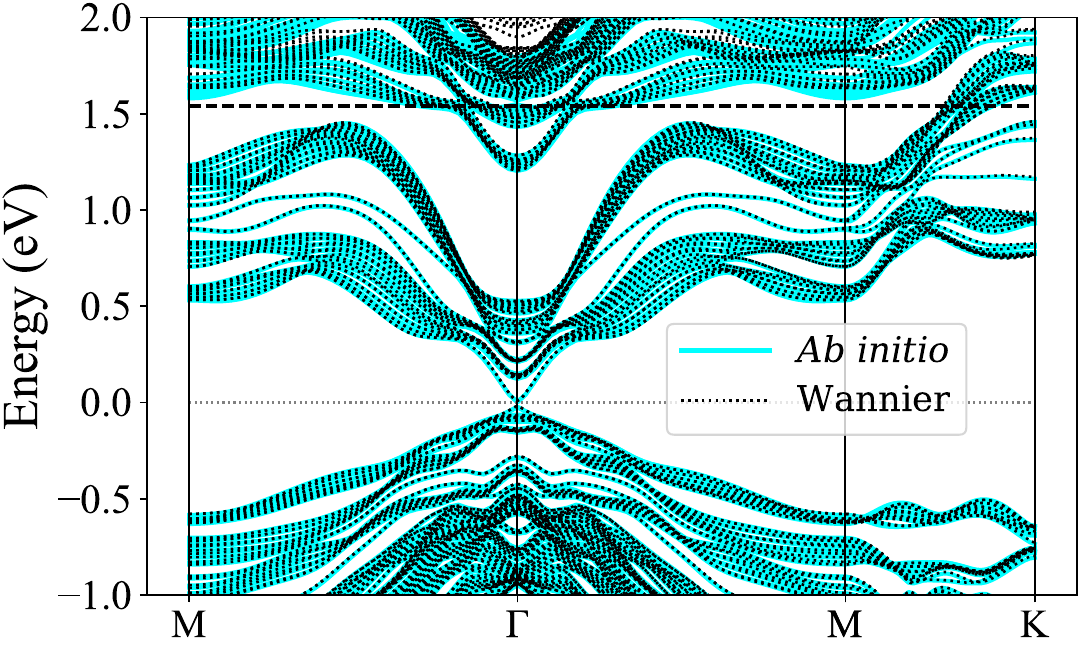}
    \caption{\textit{Ab initio} and Wannier-interpolated band structures of 11 SL MnBi$_2$Te$_4$ plotted along high symmetry path.  The Fermi energy is at 0 eV, and the dashed line at $1.54$ eV indicates the upper limit of the frozen energy window used in the Wannier disentanglement procedure.}
    \label{fig:11SL}
\end{figure}
%%%%%%%%%%%%%%%%%%%%%%%%%%%%%%%%%%%%%%%% Appendix A %%%%%%%%%%%%%%%%%%%%%%%%%%%%%%%%%%%%%%
\appendix
\section{Wannierization and Chern number details}
\label{sec:WannierDetails}
Out-of-plane lattice relaxations and self-consistent calculations were performed using a $\Gamma$-centered $\boldsymbol{k}$-grid of $16\times 16 \times
1$ with a convergence threshold of $10^{-7}$ eV, and relaxed until the forces acting on each atom were less than $10^{-2} \, \mathrm{eV}/\text{\AA}$. The non-self-consistent step in the DFT calculations was performed on a sparse $\boldsymbol{k}$-grid of $8\times 8 \times 1$, selected to balance good Wannier-interpolated band structures and efficiency. The in-plane lattice constant was fixed at $a=4.36\, \text{\AA}$ \cite{doi:10.1126/sciadv.aaz0948} for all thin film systems. In all DFT calculations, a vacuum spacing of more than 22~\AA{} was included along the out-of-plane direction to minimize spurious interactions between periodic images. All calculations in this work are performed for this A-type antiferromagnetic configuration in the absence of an external magnetic field. 

As a post-processing step, single-shot Wannier functions were constructed using the Wannier90 code package \cite{MOSTOFI20142309,Pizzi_2020}. These Wannier functions were constructed from the Te p- and Bi p-orbital character bands around the Fermi level; the details of the frozen and disentanglement energy windows used in the Wannierization procedure performed with Wannier90 are in Table~\ref{tab:wannier_windows}, as well as the Chern numbers with and without SOC in Table~\ref{tab:chern_numbers} for all thin films considered. The numbers presented were calculated using Eq.~(\ref{eq:ChernNumber}), avoiding the immediate neighborhoods of band crossings \footnotemark[\value{berryfoot}]. The results of Eq.~(\ref{eq:ChernNumber}), which suffer from band truncation, nonetheless agree within typically 1\%. For comparison with previous DFT calculations, M. M. Otrokov \textit{et al}. reported
gaps of 321 meV (1 SL), 97 meV (4 SL), and 77 meV (5 SL) \cite{PhysRevLett.122.107202}. 
We computed the response properties using the WannierBerri code package \cite{tsirkin_high_2021}, employing
a dense $\boldsymbol{k}$-grid of $1000\times 1000\times 1$ and adaptive mesh refinement with 20 iterations for Brillouin zone integration. The DFT and Wannier-interpolated band structures is shown in Fig.~\ref{fig:11SL}.

\begin{table}[h]
\centering
\begin{tabular}{c c c c}
\toprule
\multicolumn{1}{c|}{\textbf{SL}} &
\multicolumn{1}{c|}{\textbf{Inner/Outer min}} &
\multicolumn{1}{c|}{\textbf{Inner max}} &
\multicolumn{1}{c }{\textbf{Outer max}} \\
\midrule
1  & -5.05 & 1.73 & 3.16 \\
4  & -5.20 & 1.52 & 3.38 \\
5  & -5.29 & 1.54 & 3.51 \\
11 & -5.21 & 1.54 & 3.35 \\
\bottomrule
\end{tabular}
\caption{Energy windows used for Wannierization for 1, 4, 5 and 11 SL MnBi$_2$Te$_4$ thin films. All energies are given in eV relative to the Fermi energy.}
\label{tab:wannier_windows}
\end{table}

\begin{table}[h]
\centering
\begin{tabular}{c c c c c}
\toprule
\multicolumn{1}{c|}{\textbf{SOC}} &
\multicolumn{1}{c|}{\textbf{1 SL}} &
\multicolumn{1}{c|}{\textbf{4 SL}} &
\multicolumn{1}{c|}{\textbf{5 SL}} &
\multicolumn{1}{c }{\textbf{11 SL}} \\
\midrule
Yes & $-2.16\times 10^{-2}$ & $1.03\times 10^{-4}$ & $-1.01 \times 10^{0}$ & $-9.90 \times 10^{-1}$ \\
No  & $5.17\times 10^{-4}$ & $-8.67\times 10^{-6}$ & $-2.65\times 10^{-4}$ & $-1.20\times 10^{-5}$ \\
\bottomrule
\end{tabular}
\caption{Numerical results for the Chern numbers of
MnBi$_2$Te$_4$ thin films of varying thickness, 
with and without the inclusion of SOC, calculated in WannierBerri on a $300\times 300$ $\boldsymbol{k}$-grid with 
adaptive mesh refinement consisting of 60 iterations. 
}
\label{tab:chern_numbers}
\end{table}
\begin{figure*}[tb]
    \centering  \includegraphics[width=1.0\textwidth]{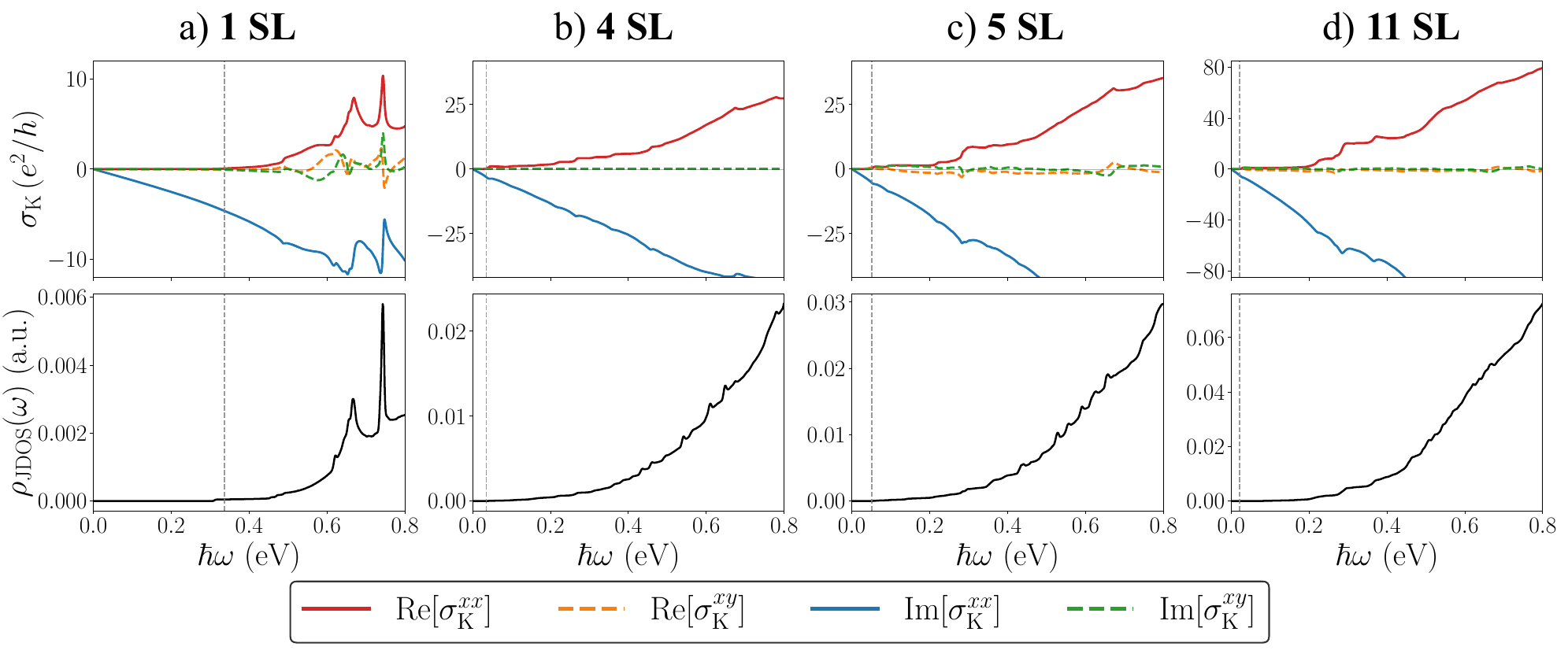}
    \caption{Kubo conductivity and joint density of states for thin films of (a) 1 SL, (b) 4 SL, (c) 5 SL, and (d) 11 SL MnBi$_2$Te$_4$. Similar to Fig.~\ref{fig:chi}(b),  the combination of parity-time symmetry in the 4 SL system, together with $C_{3z}$ rotational symmetry, enforces $\sigma_{\mathrm{K}}^{xy}(\omega)=0$  at all frequencies. The grey dashed vertical line indicates the band gap energy. }
    \label{fig:KuboCond}
\end{figure*}

\begin{figure}[t]
    \centering  \includegraphics[width=0.93\linewidth]{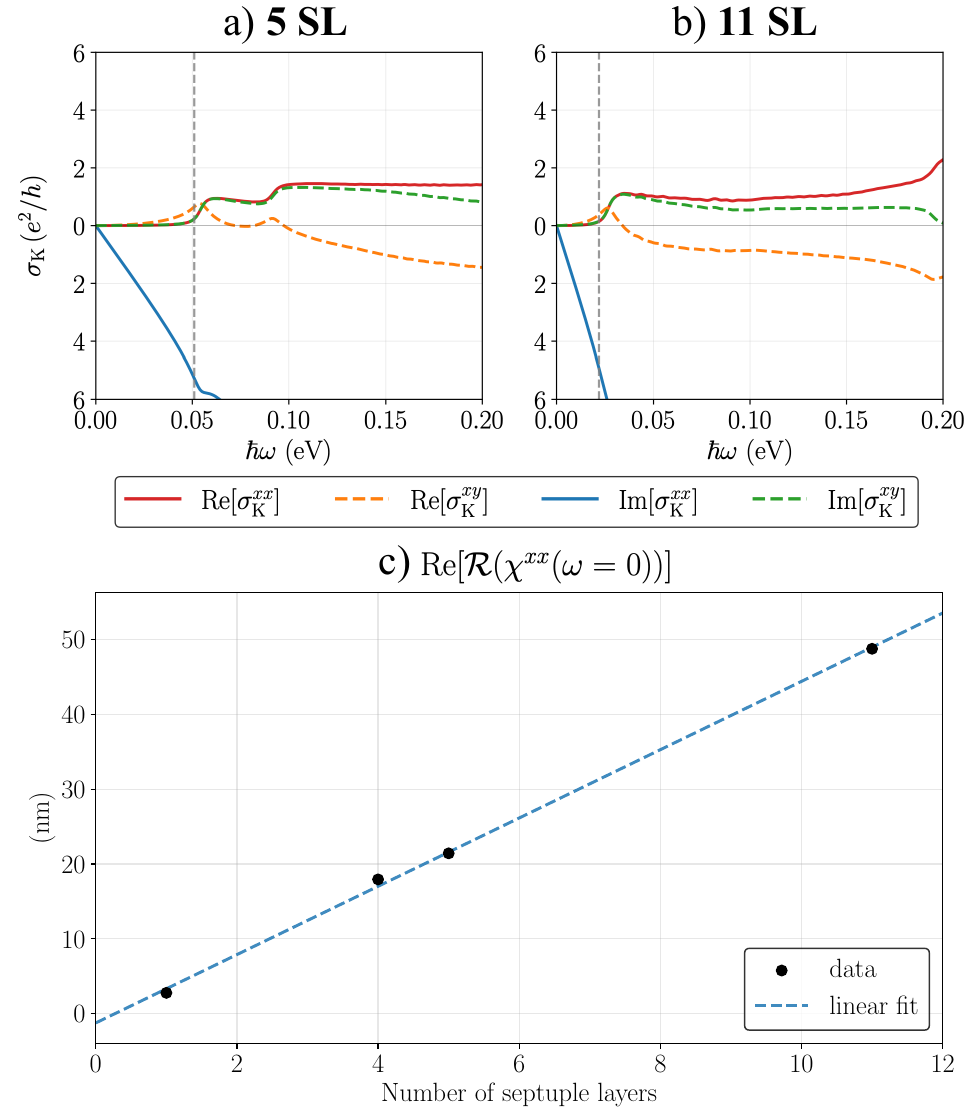}
    \caption{Low-frequency Kubo conductivity for thin films of (a) 5 SL and (b) 11 SL MnBi$_2$Te$_4$, highlighting  Re$[\sigma_\mathrm{K}^{xx}]\approx$ Im$[\sigma_\mathrm{K}^{xy}]$ within a narrow infrared energy window, which is associated with nearly perfect magnetic circular dichroism. The grey dashed vertical line indicates the band gap energy. (c) The static reactive susceptibility component $\mathcal{R}(\chi^{x x}(0))$ as a function of layers, which is well parametrized by a linear function.}
    \label{fig:MagCircDich}
\end{figure}
\section{Kubo conductivities}
\label{sec:Ku}
The Kubo conductivity tensor, defined in Eq.~(\ref{eq:Kubo}), is plotted in the upper panels of Figs.~\ref{fig:KuboCond}(a-d) as a function of photon energy in the infrared regime for 1, 4, 5 and 11 SL MnBi$_2$Te$_4$. In all systems, $\text{Re}[\sigma_\mathrm{K}^{xx}]$ exhibits the behavior of a peak followed by a plateau, as observed in other theoretical models \cite{doi:10.1126/sciadv.ado1761,PhysRevB.108.125424}. We note that our results closely reproduce the 1 SL optical conductivity reported in
B. Ghosh \textit{et al}. \cite{doi:10.1126/sciadv.ado1761} when the Hubbard parameter is adjusted from $5.34$ eV to 3 eV. For the 5 and 11 SL systems, $|\text{Re}[\sigma_\mathrm{K}^{xy}]| < e^2/h$ at low frequencies ($\hbar\omega <90$ meV and $\hbar\omega <30$ meV, respectively), indicating that the topological Hall response dominates the transverse conductivity in this frequency regime. As shown in Figs.~\ref{fig:MagCircDich}(a) and \ref{fig:MagCircDich}(b), the 5 and 11 SL Kubo conductivities satisfy Re$[\sigma_\mathrm{K}^{xx}]\approx$ Im$[\sigma_\mathrm{K}^{xy}]$ within the energy range $50 \leq \hbar\omega \leq  130$ meV and $30 \leq \hbar\omega \leq  60$ meV respectively. For comparison, the lower panels of Figs.~\ref{fig:KuboCond}(a-d) show the joint density of states (JDOS) as defined by 
\begin{equation}
    \rho_{\text{JDOS}}(\omega) = \sum_{mn} f_{nm} \int_{\mathrm{BZ}^2} \frac{d\boldsymbol{k}}{(2\pi)^2}\, \delta(E_{m\boldsymbol{k}}-E_{n\boldsymbol{k}}-\hbar \omega).
\end{equation}

\section{Layer-resolved reactive susceptibility}
\label{sec:layer_resolved}
Here, we consider the static limit of the reactive susceptibility component given in Eq.~(\ref{eq:reactive_quantum_metric}) and decompose it into intralayer and interlayer contributions, following a procedure similar to that recently developed for the quantum anomalous Hall conductivity \cite{PhysRevB.98.245117,PhysRevB.84.241106,PhysRevB.95.121114,li2025high}. Since the Wannier functions for $N_L$-SL MnBi$_2$Te$_4$ are obtained through single-shot Wannierization, each Wannier center is associated with a particular layer $\ell \in \{1,\ldots,N_L\}$. The unitary transformation between Bloch states in the Wannier and Bloch frames is then
\begin{equation}
    u_{\alpha_{\ell} \boldsymbol{k}}(\boldsymbol{x}) = \sum_{n}U_{n\alpha_{\ell}}(\boldsymbol{k}) u_{n\boldsymbol{k}}(\boldsymbol{x}),
\end{equation}
where $\alpha_\ell$ labels a Wannier orbital index $\alpha$ belonging to layer $\ell$. We begin by recasting  Eq.~(\ref{eq:reactive_quantum_metric}) as
\begin{equation}
    \mathcal{R}(\chi^{x x}(0)) = 2e^2  \sum_{cv} \int_{\mathrm{BZ}^2} \frac{d\boldsymbol{k}}{(2\pi)^2} 
\frac{\xi^{x}_{vc}(\boldsymbol{k}) \xi^x_{cv}(\boldsymbol{k})}
{E_{c\boldsymbol{k}} - E_{v\boldsymbol{k}}},
\end{equation}
where the indices $v$ and $c$ label valence and conduction bands, respectively. Since $E_{c\boldsymbol{k}}>E_{v\boldsymbol{k}}$, we can write 
\begin{equation}
    \frac{\xi^{x}_{vc}(\boldsymbol{k}) \xi^x_{cv}(\boldsymbol{k})}
{E_{c\boldsymbol{k}} - E_{v\boldsymbol{k}}}=\frac{\xi^{x}_{vc}(\boldsymbol{k})}
{\sqrt{E_{c\boldsymbol{k}} - E_{v\boldsymbol{k}}}} \frac{\xi^x_{cv}(\boldsymbol{k})}
{\sqrt{E_{c\boldsymbol{k}} - E_{v\boldsymbol{k}}}}.
\end{equation}
We follow a construction similar to that of Varnava and Vanderbilt \cite{PhysRevB.98.245117}. However, because our Wannier functions are a linear combination of Bloch states from both occupied valence bands and a subset of unoccupied conduction bands, it is necessary to 
 resolve both the occupied and unoccupied subspaces
 \begin{equation}
    \frac{\xi^{x}_{vc}(\boldsymbol{k}) \xi^x_{cv}(\boldsymbol{k})}
{E_{c\boldsymbol{k}} - E_{v\boldsymbol{k}}}=\sum_{c'v'}\frac{\xi^{x}_{vc}(\boldsymbol{k})}
{\sqrt{E_{c\boldsymbol{k}} - E_{v\boldsymbol{k}}}} \frac{\xi^x_{c'v'}(\boldsymbol{k})}
{\sqrt{E_{c'\boldsymbol{k}} - E_{v'\boldsymbol{k}}}}\delta_{cc'}\delta_{vv'}.
\end{equation}
Define the layer-resolved quantity
\begin{equation}
    \rho_{nm}(\ell,\boldsymbol{k})= \sum_{\alpha_\ell} U_{n\alpha_{\ell}}U^*_{m\alpha_\ell},
\end{equation}
which satisfies 
\begin{equation}
    \sum_{\ell=1}^{N_{L}} \rho_{nm}(\ell,\boldsymbol{k})=\delta_{nm}.
\end{equation}
It follows that
\begin{equation}
\begin{aligned}
    \frac{\xi^{x}_{vc}(\boldsymbol{k})\, \xi^x_{cv}(\boldsymbol{k})}
         {E_{c\boldsymbol{k}} - E_{v\boldsymbol{k}}}
    ={}& \sum_{\ell=1}^{N_{L}} \sum_{\ell'=1}^{N_{L}} \sum_{c'v'}
      \frac{\xi^{x}_{vc}(\boldsymbol{k})}{\sqrt{E_{c\boldsymbol{k}} - E_{v\boldsymbol{k}}}} \\
    &\times \frac{\xi^x_{c'v'}(\boldsymbol{k})}{\sqrt{E_{c'\boldsymbol{k}} - E_{v'\boldsymbol{k}}}}\,
      \rho_{cc'}(\ell',\boldsymbol{k})\, \rho_{vv'}(\ell,\boldsymbol{k}),
\end{aligned}
\end{equation}
and so the static reactive susceptibility becomes
\begin{equation}
    \label{eq:Layer_reactive_chi}
    \mathcal{R}(\chi^{x x}(0)) =   \sum_{\ell,\ell'=1}^{N_{L}} \mathcal{R}(\chi^{x x}_{\ell\ell'}(0)),
\end{equation}
where we have defined 
\begin{equation}
\begin{aligned}
    \mathcal{R}(\chi^{x x}_{\ell\ell'}(0)) &= 2e^2  \int_{\mathrm{BZ}^2} \frac{d\boldsymbol{k}}{(2\pi)^2}  \sum_{cc'vv'}
      \frac{\xi^{x}_{vc}(\boldsymbol{k})\xi^x_{c'v'}(\boldsymbol{k})}{\sqrt{(E_{c\boldsymbol{k}} - E_{v\boldsymbol{k}})(E_{c'\boldsymbol{k}} - E_{v'\boldsymbol{k}})}} \\
    &\qquad\times
      \rho_{cc'}(\ell',\boldsymbol{k})\, \rho_{vv'}(\ell,\boldsymbol{k}).
\end{aligned}
\end{equation}
Eq.~(\ref{eq:Layer_reactive_chi}) naturally decomposes into intralayer contributions, $\mathcal{R}(\chi^{x x}_{\ell\ell}(0))$, and interlayer contributions, $\mathcal{R}(\chi^{x x}_{\ell\ell'}(0))$ with $\ell\neq \ell'$.

%\textcolor{blue}{Note: convention in WannierBerri for $U(\boldsymbol{k})$ differ in that $U^{\text{WB}}_{\alpha n}=U^*_{n\alpha}$.}

%%%%%%%%%%%%%%%%%%%%%%%%%%%%%%%%%%%%%%% Bibliography %%%%%%%%%%%%%%%%%%%%%%%%%%%%%%%%%%%%%

%\newpage
\clearpage
%\bibliographystyle{apsrev4-2}
%\bibliography{Bibliography}

\begin{thebibliography}{68}%
\makeatletter
\providecommand \@ifxundefined [1]{%
 \@ifx{#1\undefined}
}%
\providecommand \@ifnum [1]{%
 \ifnum #1\expandafter \@firstoftwo
 \else \expandafter \@secondoftwo
 \fi
}%
\providecommand \@ifx [1]{%
 \ifx #1\expandafter \@firstoftwo
 \else \expandafter \@secondoftwo
 \fi
}%
\providecommand \natexlab [1]{#1}%
\providecommand \enquote  [1]{``#1''}%
\providecommand \bibnamefont  [1]{#1}%
\providecommand \bibfnamefont [1]{#1}%
\providecommand \citenamefont [1]{#1}%
\providecommand \href@noop [0]{\@secondoftwo}%
\providecommand \href [0]{\begingroup \@sanitize@url \@href}%
\providecommand \@href[1]{\@@startlink{#1}\@@href}%
\providecommand \@@href[1]{\endgroup#1\@@endlink}%
\providecommand \@sanitize@url [0]{\catcode `\\12\catcode `\$12\catcode `\&12\catcode `\#12\catcode `\^12\catcode `\_12\catcode `\%12\relax}%
\providecommand \@@startlink[1]{}%
\providecommand \@@endlink[0]{}%
\providecommand \url  [0]{\begingroup\@sanitize@url \@url }%
\providecommand \@url [1]{\endgroup\@href {#1}{\urlprefix }}%
\providecommand \urlprefix  [0]{URL }%
\providecommand \Eprint [0]{\href }%
\providecommand \doibase [0]{https://doi.org/}%
\providecommand \selectlanguage [0]{\@gobble}%
\providecommand \bibinfo  [0]{\@secondoftwo}%
\providecommand \bibfield  [0]{\@secondoftwo}%
\providecommand \translation [1]{[#1]}%
\providecommand \BibitemOpen [0]{}%
\providecommand \bibitemStop [0]{}%
\providecommand \bibitemNoStop [0]{.\EOS\space}%
\providecommand \EOS [0]{\spacefactor3000\relax}%
\providecommand \BibitemShut  [1]{\csname bibitem#1\endcsname}%
\let\auto@bib@innerbib\@empty
%</preamble>
\bibitem [{\citenamefont {Klitzing}\ \emph {et~al.}(1980)\citenamefont {Klitzing}, \citenamefont {Dorda},\ and\ \citenamefont {Pepper}}]{PhysRevLett.45.494}%
  \BibitemOpen
  \bibfield  {author} {\bibinfo {author} {\bibfnamefont {K.~v.}\ \bibnamefont {Klitzing}}, \bibinfo {author} {\bibfnamefont {G.}~\bibnamefont {Dorda}},\ and\ \bibinfo {author} {\bibfnamefont {M.}~\bibnamefont {Pepper}},\ }\bibfield  {title} {\bibinfo {title} {New method for high-accuracy determination of the fine-structure constant based on quantized hall resistance},\ }\href {https://doi.org/10.1103/PhysRevLett.45.494} {\bibfield  {journal} {\bibinfo  {journal} {Phys. Rev. Lett.}\ }\textbf {\bibinfo {volume} {45}},\ \bibinfo {pages} {494} (\bibinfo {year} {1980})}\BibitemShut {NoStop}%
\bibitem [{\citenamefont {Jain}(2007)}]{Jain_2007}%
  \BibitemOpen
  \bibfield  {author} {\bibinfo {author} {\bibfnamefont {J.~K.}\ \bibnamefont {Jain}},\ }\href@noop {} {\emph {\bibinfo {title} {Composite Fermions}}}\ (\bibinfo  {publisher} {Cambridge University Press},\ \bibinfo {year} {2007})\BibitemShut {NoStop}%
\bibitem [{\citenamefont {Haldane}(1988)}]{PhysRevLett.61.2015}%
  \BibitemOpen
  \bibfield  {author} {\bibinfo {author} {\bibfnamefont {F.~D.~M.}\ \bibnamefont {Haldane}},\ }\bibfield  {title} {\bibinfo {title} {Model for a quantum hall effect without landau levels: Condensed-matter realization of the ``parity anomaly"},\ }\href {https://doi.org/10.1103/PhysRevLett.61.2015} {\bibfield  {journal} {\bibinfo  {journal} {Phys. Rev. Lett.}\ }\textbf {\bibinfo {volume} {61}},\ \bibinfo {pages} {2015} (\bibinfo {year} {1988})}\BibitemShut {NoStop}%
\bibitem [{\citenamefont {Hasan}\ and\ \citenamefont {Kane}(2010)}]{RevModPhys.82.3045}%
  \BibitemOpen
  \bibfield  {author} {\bibinfo {author} {\bibfnamefont {M.~Z.}\ \bibnamefont {Hasan}}\ and\ \bibinfo {author} {\bibfnamefont {C.~L.}\ \bibnamefont {Kane}},\ }\bibfield  {title} {\bibinfo {title} {Colloquium: Topological insulators},\ }\href {https://doi.org/10.1103/RevModPhys.82.3045} {\bibfield  {journal} {\bibinfo  {journal} {Rev. Mod. Phys.}\ }\textbf {\bibinfo {volume} {82}},\ \bibinfo {pages} {3045} (\bibinfo {year} {2010})}\BibitemShut {NoStop}%
\bibitem [{\citenamefont {Bosnar}\ \emph {et~al.}(2023)\citenamefont {Bosnar}, \citenamefont {Vyazovskaya}, \citenamefont {Petrov}, \citenamefont {Chulkov},\ and\ \citenamefont {Otrokov}}]{bosnar_high_2023}%
  \BibitemOpen
  \bibfield  {author} {\bibinfo {author} {\bibfnamefont {M.}~\bibnamefont {Bosnar}}, \bibinfo {author} {\bibfnamefont {A.~Y.}\ \bibnamefont {Vyazovskaya}}, \bibinfo {author} {\bibfnamefont {E.~K.}\ \bibnamefont {Petrov}}, \bibinfo {author} {\bibfnamefont {E.~V.}\ \bibnamefont {Chulkov}},\ and\ \bibinfo {author} {\bibfnamefont {M.~M.}\ \bibnamefont {Otrokov}},\ }\bibfield  {title} {\bibinfo {title} {High {Chern} number van der {Waals} magnetic topological multilayers {{MnBi\textsubscript{2}Te\textsubscript{4}}}/{hBN}},\ }\href {https://doi.org/10.1038/s41699-023-00396-y} {\bibfield  {journal} {\bibinfo  {journal} {npj 2D Materials and Applications}\ }\textbf {\bibinfo {volume} {7}},\ \bibinfo {pages} {33} (\bibinfo {year} {2023})}\BibitemShut {NoStop}%
\bibitem [{\citenamefont {Li}\ \emph {et~al.}(2025{\natexlab{a}})\citenamefont {Li}, \citenamefont {Wu},\ and\ \citenamefont {Weng}}]{li_high-chern-number_2025}%
  \BibitemOpen
  \bibfield  {author} {\bibinfo {author} {\bibfnamefont {J.}~\bibnamefont {Li}}, \bibinfo {author} {\bibfnamefont {Q.}~\bibnamefont {Wu}},\ and\ \bibinfo {author} {\bibfnamefont {H.}~\bibnamefont {Weng}},\ }\bibfield  {title} {\bibinfo {title} {High-{Chern}-number {Quantum} anomalous {Hall} insulators in mixing-stacked {{MnBi\textsubscript{2}Te\textsubscript{4}}} thin films},\ }\href {https://doi.org/10.1038/s41535-025-00775-2} {\bibfield  {journal} {\bibinfo  {journal} {npj Quantum Materials}\ }\textbf {\bibinfo {volume} {10}},\ \bibinfo {pages} {53} (\bibinfo {year} {2025}{\natexlab{a}})},\ \bibinfo {note} {publisher: Nature Publishing Group}\BibitemShut {NoStop}%
\bibitem [{\citenamefont {Sattar}\ \emph {et~al.}(2025)\citenamefont {Sattar}, \citenamefont {Hedman},\ and\ \citenamefont {Canali}}]{PhysRevResearch.7.023024}%
  \BibitemOpen
  \bibfield  {author} {\bibinfo {author} {\bibfnamefont {S.}~\bibnamefont {Sattar}}, \bibinfo {author} {\bibfnamefont {D.}~\bibnamefont {Hedman}},\ and\ \bibinfo {author} {\bibfnamefont {C.~M.}\ \bibnamefont {Canali}},\ }\bibfield  {title} {\bibinfo {title} {Surface reconstructions in thin films of magnetic topological insulator {MnBi\textsubscript{2}Te\textsubscript{4}}},\ }\href {https://doi.org/10.1103/PhysRevResearch.7.023024} {\bibfield  {journal} {\bibinfo  {journal} {Phys. Rev. Res.}\ }\textbf {\bibinfo {volume} {7}},\ \bibinfo {pages} {023024} (\bibinfo {year} {2025})}\BibitemShut {NoStop}%
\bibitem [{\citenamefont {Ou}\ \emph {et~al.}(2018)\citenamefont {Ou}, \citenamefont {Liu}, \citenamefont {Jiang}, \citenamefont {Feng}, \citenamefont {Zhao}, \citenamefont {Wu}, \citenamefont {Wang}, \citenamefont {Li}, \citenamefont {Song}, \citenamefont {Wang}, \citenamefont {Wang}, \citenamefont {Wu}, \citenamefont {Wang}, \citenamefont {He}, \citenamefont {Ma},\ and\ \citenamefont {Xue}}]{https://doi.org/10.1002/adma.201703062}%
  \BibitemOpen
  \bibfield  {author} {\bibinfo {author} {\bibfnamefont {Y.}~\bibnamefont {Ou}}, \bibinfo {author} {\bibfnamefont {C.}~\bibnamefont {Liu}}, \bibinfo {author} {\bibfnamefont {G.}~\bibnamefont {Jiang}}, \bibinfo {author} {\bibfnamefont {Y.}~\bibnamefont {Feng}}, \bibinfo {author} {\bibfnamefont {D.}~\bibnamefont {Zhao}}, \bibinfo {author} {\bibfnamefont {W.}~\bibnamefont {Wu}}, \bibinfo {author} {\bibfnamefont {X.-X.}\ \bibnamefont {Wang}}, \bibinfo {author} {\bibfnamefont {W.}~\bibnamefont {Li}}, \bibinfo {author} {\bibfnamefont {C.}~\bibnamefont {Song}}, \bibinfo {author} {\bibfnamefont {L.-L.}\ \bibnamefont {Wang}}, \bibinfo {author} {\bibfnamefont {W.}~\bibnamefont {Wang}}, \bibinfo {author} {\bibfnamefont {W.}~\bibnamefont {Wu}}, \bibinfo {author} {\bibfnamefont {Y.}~\bibnamefont {Wang}}, \bibinfo {author} {\bibfnamefont {K.}~\bibnamefont {He}}, \bibinfo {author} {\bibfnamefont {X.-C.}\ \bibnamefont {Ma}},\ and\ \bibinfo {author} {\bibfnamefont {Q.-K.}\ \bibnamefont {Xue}},\ }\bibfield  {title} {\bibinfo
  {title} {Enhancing the quantum anomalous hall effect by magnetic codoping in a topological insulator},\ }\href {https://doi.org/https://doi.org/10.1002/adma.201703062} {\bibfield  {journal} {\bibinfo  {journal} {Advanced Materials}\ }\textbf {\bibinfo {volume} {30}},\ \bibinfo {pages} {1703062} (\bibinfo {year} {2018})}\BibitemShut {NoStop}%
\bibitem [{\citenamefont {Chang}\ \emph {et~al.}(2013)\citenamefont {Chang}, \citenamefont {Zhang}, \citenamefont {Feng}, \citenamefont {Shen}, \citenamefont {Zhang}, \citenamefont {Guo}, \citenamefont {Li}, \citenamefont {Ou}, \citenamefont {Wei}, \citenamefont {Wang}, \citenamefont {Ji}, \citenamefont {Feng}, \citenamefont {Ji}, \citenamefont {Chen}, \citenamefont {Jia}, \citenamefont {Dai}, \citenamefont {Fang}, \citenamefont {Zhang}, \citenamefont {He}, \citenamefont {Wang}, \citenamefont {Lu}, \citenamefont {Ma},\ and\ \citenamefont {Xue}}]{doi:10.1126/science.1234414}%
  \BibitemOpen
  \bibfield  {author} {\bibinfo {author} {\bibfnamefont {C.-Z.}\ \bibnamefont {Chang}}, \bibinfo {author} {\bibfnamefont {J.}~\bibnamefont {Zhang}}, \bibinfo {author} {\bibfnamefont {X.}~\bibnamefont {Feng}}, \bibinfo {author} {\bibfnamefont {J.}~\bibnamefont {Shen}}, \bibinfo {author} {\bibfnamefont {Z.}~\bibnamefont {Zhang}}, \bibinfo {author} {\bibfnamefont {M.}~\bibnamefont {Guo}}, \bibinfo {author} {\bibfnamefont {K.}~\bibnamefont {Li}}, \bibinfo {author} {\bibfnamefont {Y.}~\bibnamefont {Ou}}, \bibinfo {author} {\bibfnamefont {P.}~\bibnamefont {Wei}}, \bibinfo {author} {\bibfnamefont {L.-L.}\ \bibnamefont {Wang}}, \bibinfo {author} {\bibfnamefont {Z.-Q.}\ \bibnamefont {Ji}}, \bibinfo {author} {\bibfnamefont {Y.}~\bibnamefont {Feng}}, \bibinfo {author} {\bibfnamefont {S.}~\bibnamefont {Ji}}, \bibinfo {author} {\bibfnamefont {X.}~\bibnamefont {Chen}}, \bibinfo {author} {\bibfnamefont {J.}~\bibnamefont {Jia}}, \bibinfo {author} {\bibfnamefont {X.}~\bibnamefont {Dai}}, \bibinfo {author} {\bibfnamefont
  {Z.}~\bibnamefont {Fang}}, \bibinfo {author} {\bibfnamefont {S.-C.}\ \bibnamefont {Zhang}}, \bibinfo {author} {\bibfnamefont {K.}~\bibnamefont {He}}, \bibinfo {author} {\bibfnamefont {Y.}~\bibnamefont {Wang}}, \bibinfo {author} {\bibfnamefont {L.}~\bibnamefont {Lu}}, \bibinfo {author} {\bibfnamefont {X.-C.}\ \bibnamefont {Ma}},\ and\ \bibinfo {author} {\bibfnamefont {Q.-K.}\ \bibnamefont {Xue}},\ }\bibfield  {title} {\bibinfo {title} {Experimental observation of the quantum anomalous hall effect in a magnetic topological insulator},\ }\href {https://doi.org/10.1126/science.1234414} {\bibfield  {journal} {\bibinfo  {journal} {Science}\ }\textbf {\bibinfo {volume} {340}},\ \bibinfo {pages} {167} (\bibinfo {year} {2013})}\BibitemShut {NoStop}%
\bibitem [{\citenamefont {Kou}\ \emph {et~al.}(2014)\citenamefont {Kou}, \citenamefont {Guo}, \citenamefont {Fan}, \citenamefont {Pan}, \citenamefont {Lang}, \citenamefont {Jiang}, \citenamefont {Shao}, \citenamefont {Nie}, \citenamefont {Murata}, \citenamefont {Tang}, \citenamefont {Wang}, \citenamefont {He}, \citenamefont {Lee}, \citenamefont {Lee},\ and\ \citenamefont {Wang}}]{PhysRevLett.113.137201}%
  \BibitemOpen
  \bibfield  {author} {\bibinfo {author} {\bibfnamefont {X.}~\bibnamefont {Kou}}, \bibinfo {author} {\bibfnamefont {S.-T.}\ \bibnamefont {Guo}}, \bibinfo {author} {\bibfnamefont {Y.}~\bibnamefont {Fan}}, \bibinfo {author} {\bibfnamefont {L.}~\bibnamefont {Pan}}, \bibinfo {author} {\bibfnamefont {M.}~\bibnamefont {Lang}}, \bibinfo {author} {\bibfnamefont {Y.}~\bibnamefont {Jiang}}, \bibinfo {author} {\bibfnamefont {Q.}~\bibnamefont {Shao}}, \bibinfo {author} {\bibfnamefont {T.}~\bibnamefont {Nie}}, \bibinfo {author} {\bibfnamefont {K.}~\bibnamefont {Murata}}, \bibinfo {author} {\bibfnamefont {J.}~\bibnamefont {Tang}}, \bibinfo {author} {\bibfnamefont {Y.}~\bibnamefont {Wang}}, \bibinfo {author} {\bibfnamefont {L.}~\bibnamefont {He}}, \bibinfo {author} {\bibfnamefont {T.-K.}\ \bibnamefont {Lee}}, \bibinfo {author} {\bibfnamefont {W.-L.}\ \bibnamefont {Lee}},\ and\ \bibinfo {author} {\bibfnamefont {K.~L.}\ \bibnamefont {Wang}},\ }\bibfield  {title} {\bibinfo {title} {Scale-invariant quantum anomalous hall effect
  in magnetic topological insulators beyond the two-dimensional limit},\ }\href {https://doi.org/10.1103/PhysRevLett.113.137201} {\bibfield  {journal} {\bibinfo  {journal} {Phys. Rev. Lett.}\ }\textbf {\bibinfo {volume} {113}},\ \bibinfo {pages} {137201} (\bibinfo {year} {2014})}\BibitemShut {NoStop}%
\bibitem [{\citenamefont {Vanderbilt}(2018)}]{vanderbilt2018berry}%
  \BibitemOpen
  \bibfield  {author} {\bibinfo {author} {\bibfnamefont {D.}~\bibnamefont {Vanderbilt}},\ }\href@noop {} {\emph {\bibinfo {title} {Berry phases in electronic structure theory: electric polarization, orbital magnetization and topological insulators}}}\ (\bibinfo  {publisher} {Cambridge University Press},\ \bibinfo {year} {2018})\BibitemShut {NoStop}%
\bibitem [{\citenamefont {Deng}\ \emph {et~al.}(2020)\citenamefont {Deng}, \citenamefont {Yu}, \citenamefont {Shi}, \citenamefont {Guo}, \citenamefont {Xu}, \citenamefont {Wang}, \citenamefont {Chen},\ and\ \citenamefont {Zhang}}]{doi:10.1126/science.aax8156}%
  \BibitemOpen
  \bibfield  {author} {\bibinfo {author} {\bibfnamefont {Y.}~\bibnamefont {Deng}}, \bibinfo {author} {\bibfnamefont {Y.}~\bibnamefont {Yu}}, \bibinfo {author} {\bibfnamefont {M.~Z.}\ \bibnamefont {Shi}}, \bibinfo {author} {\bibfnamefont {Z.}~\bibnamefont {Guo}}, \bibinfo {author} {\bibfnamefont {Z.}~\bibnamefont {Xu}}, \bibinfo {author} {\bibfnamefont {J.}~\bibnamefont {Wang}}, \bibinfo {author} {\bibfnamefont {X.~H.}\ \bibnamefont {Chen}},\ and\ \bibinfo {author} {\bibfnamefont {Y.}~\bibnamefont {Zhang}},\ }\bibfield  {title} {\bibinfo {title} {Quantum anomalous hall effect in intrinsic magnetic topological insulator {MnBi\textsubscript{2}Te\textsubscript{4}}},\ }\href {https://doi.org/10.1126/science.aax8156} {\bibfield  {journal} {\bibinfo  {journal} {Science}\ }\textbf {\bibinfo {volume} {367}},\ \bibinfo {pages} {895} (\bibinfo {year} {2020})}\BibitemShut {NoStop}%
\bibitem [{\citenamefont {Otrokov}\ \emph {et~al.}(2019{\natexlab{a}})\citenamefont {Otrokov}, \citenamefont {Rusinov}, \citenamefont {Blanco-Rey}, \citenamefont {Hoffmann}, \citenamefont {Vyazovskaya}, \citenamefont {Eremeev}, \citenamefont {Ernst}, \citenamefont {Echenique}, \citenamefont {Arnau},\ and\ \citenamefont {Chulkov}}]{PhysRevLett.122.107202}%
  \BibitemOpen
  \bibfield  {author} {\bibinfo {author} {\bibfnamefont {M.~M.}\ \bibnamefont {Otrokov}}, \bibinfo {author} {\bibfnamefont {I.~P.}\ \bibnamefont {Rusinov}}, \bibinfo {author} {\bibfnamefont {M.}~\bibnamefont {Blanco-Rey}}, \bibinfo {author} {\bibfnamefont {M.}~\bibnamefont {Hoffmann}}, \bibinfo {author} {\bibfnamefont {A.~Y.}\ \bibnamefont {Vyazovskaya}}, \bibinfo {author} {\bibfnamefont {S.~V.}\ \bibnamefont {Eremeev}}, \bibinfo {author} {\bibfnamefont {A.}~\bibnamefont {Ernst}}, \bibinfo {author} {\bibfnamefont {P.~M.}\ \bibnamefont {Echenique}}, \bibinfo {author} {\bibfnamefont {A.}~\bibnamefont {Arnau}},\ and\ \bibinfo {author} {\bibfnamefont {E.~V.}\ \bibnamefont {Chulkov}},\ }\bibfield  {title} {\bibinfo {title} {Unique thickness-dependent properties of the van der waals interlayer antiferromagnet {MnBi\textsubscript{2}Te\textsubscript{4}} films},\ }\href {https://doi.org/10.1103/PhysRevLett.122.107202} {\bibfield  {journal} {\bibinfo  {journal} {Phys. Rev. Lett.}\ }\textbf {\bibinfo {volume} {122}},\
  \bibinfo {pages} {107202} (\bibinfo {year} {2019}{\natexlab{a}})}\BibitemShut {NoStop}%
\bibitem [{\citenamefont {Lei}\ and\ \citenamefont {MacDonald}(2023)}]{PhysRevB.108.125424}%
  \BibitemOpen
  \bibfield  {author} {\bibinfo {author} {\bibfnamefont {C.}~\bibnamefont {Lei}}\ and\ \bibinfo {author} {\bibfnamefont {A.~H.}\ \bibnamefont {MacDonald}},\ }\bibfield  {title} {\bibinfo {title} {Kerr, faraday, and magnetoelectric effects in {MnBi\textsubscript{2}Te\textsubscript{4}} thin films},\ }\href {https://doi.org/10.1103/PhysRevB.108.125424} {\bibfield  {journal} {\bibinfo  {journal} {Phys. Rev. B}\ }\textbf {\bibinfo {volume} {108}},\ \bibinfo {pages} {125424} (\bibinfo {year} {2023})}\BibitemShut {NoStop}%
\bibitem [{\citenamefont {Liu}\ \emph {et~al.}(2020)\citenamefont {Liu}, \citenamefont {Wang}, \citenamefont {Li}, \citenamefont {Wu}, \citenamefont {Li}, \citenamefont {Li}, \citenamefont {He}, \citenamefont {Xu}, \citenamefont {Zhang},\ and\ \citenamefont {Wang}}]{liu2020robust}%
  \BibitemOpen
  \bibfield  {author} {\bibinfo {author} {\bibfnamefont {C.}~\bibnamefont {Liu}}, \bibinfo {author} {\bibfnamefont {Y.}~\bibnamefont {Wang}}, \bibinfo {author} {\bibfnamefont {H.}~\bibnamefont {Li}}, \bibinfo {author} {\bibfnamefont {Y.}~\bibnamefont {Wu}}, \bibinfo {author} {\bibfnamefont {Y.}~\bibnamefont {Li}}, \bibinfo {author} {\bibfnamefont {J.}~\bibnamefont {Li}}, \bibinfo {author} {\bibfnamefont {K.}~\bibnamefont {He}}, \bibinfo {author} {\bibfnamefont {Y.}~\bibnamefont {Xu}}, \bibinfo {author} {\bibfnamefont {J.}~\bibnamefont {Zhang}},\ and\ \bibinfo {author} {\bibfnamefont {Y.}~\bibnamefont {Wang}},\ }\bibfield  {title} {\bibinfo {title} {Robust axion insulator and {Chern} insulator phases in a two-dimensional antiferromagnetic topological insulator},\ }\href@noop {} {\bibfield  {journal} {\bibinfo  {journal} {Nature materials}\ }\textbf {\bibinfo {volume} {19}},\ \bibinfo {pages} {522} (\bibinfo {year} {2020})}\BibitemShut {NoStop}%
\bibitem [{\citenamefont {Zhang}\ \emph {et~al.}(2019)\citenamefont {Zhang}, \citenamefont {Shi}, \citenamefont {Zhu}, \citenamefont {Xing}, \citenamefont {Zhang},\ and\ \citenamefont {Wang}}]{PhysRevLett.122.206401}%
  \BibitemOpen
  \bibfield  {author} {\bibinfo {author} {\bibfnamefont {D.}~\bibnamefont {Zhang}}, \bibinfo {author} {\bibfnamefont {M.}~\bibnamefont {Shi}}, \bibinfo {author} {\bibfnamefont {T.}~\bibnamefont {Zhu}}, \bibinfo {author} {\bibfnamefont {D.}~\bibnamefont {Xing}}, \bibinfo {author} {\bibfnamefont {H.}~\bibnamefont {Zhang}},\ and\ \bibinfo {author} {\bibfnamefont {J.}~\bibnamefont {Wang}},\ }\bibfield  {title} {\bibinfo {title} {Topological axion states in the magnetic insulator {MnBi\textsubscript{2}Te\textsubscript{4}} with the quantized magnetoelectric effect},\ }\href {https://doi.org/10.1103/PhysRevLett.122.206401} {\bibfield  {journal} {\bibinfo  {journal} {Phys. Rev. Lett.}\ }\textbf {\bibinfo {volume} {122}},\ \bibinfo {pages} {206401} (\bibinfo {year} {2019})}\BibitemShut {NoStop}%
\bibitem [{\citenamefont {Otrokov}\ \emph {et~al.}(2019{\natexlab{b}})\citenamefont {Otrokov}, \citenamefont {Klimovskikh}, \citenamefont {Bentmann}, \citenamefont {Estyunin}, \citenamefont {Zeugner}, \citenamefont {Aliev}, \citenamefont {Ga{\ss}}, \citenamefont {Wolter}, \citenamefont {Koroleva}, \citenamefont {Shikin} \emph {et~al.}}]{otrokov2019prediction}%
  \BibitemOpen
  \bibfield  {author} {\bibinfo {author} {\bibfnamefont {M.~M.}\ \bibnamefont {Otrokov}}, \bibinfo {author} {\bibfnamefont {I.~I.}\ \bibnamefont {Klimovskikh}}, \bibinfo {author} {\bibfnamefont {H.}~\bibnamefont {Bentmann}}, \bibinfo {author} {\bibfnamefont {D.}~\bibnamefont {Estyunin}}, \bibinfo {author} {\bibfnamefont {A.}~\bibnamefont {Zeugner}}, \bibinfo {author} {\bibfnamefont {Z.~S.}\ \bibnamefont {Aliev}}, \bibinfo {author} {\bibfnamefont {S.}~\bibnamefont {Ga{\ss}}}, \bibinfo {author} {\bibfnamefont {A.}~\bibnamefont {Wolter}}, \bibinfo {author} {\bibfnamefont {A.}~\bibnamefont {Koroleva}}, \bibinfo {author} {\bibfnamefont {A.~M.}\ \bibnamefont {Shikin}}, \emph {et~al.},\ }\bibfield  {title} {\bibinfo {title} {Prediction and observation of an antiferromagnetic topological insulator},\ }\href@noop {} {\bibfield  {journal} {\bibinfo  {journal} {Nature}\ }\textbf {\bibinfo {volume} {576}},\ \bibinfo {pages} {416} (\bibinfo {year} {2019}{\natexlab{b}})}\BibitemShut {NoStop}%
\bibitem [{\citenamefont {Liang}\ \emph {et~al.}(2025)\citenamefont {Liang}, \citenamefont {Li}, \citenamefont {An}, \citenamefont {Ren}, \citenamefont {Qiao},\ and\ \citenamefont {Niu}}]{PhysRevLett.134.116603}%
  \BibitemOpen
  \bibfield  {author} {\bibinfo {author} {\bibfnamefont {W.}~\bibnamefont {Liang}}, \bibinfo {author} {\bibfnamefont {Z.}~\bibnamefont {Li}}, \bibinfo {author} {\bibfnamefont {J.}~\bibnamefont {An}}, \bibinfo {author} {\bibfnamefont {Y.}~\bibnamefont {Ren}}, \bibinfo {author} {\bibfnamefont {Z.}~\bibnamefont {Qiao}},\ and\ \bibinfo {author} {\bibfnamefont {Q.}~\bibnamefont {Niu}},\ }\bibfield  {title} {\bibinfo {title} {Chern number tunable quantum anomalous hall effect in compensated antiferromagnets},\ }\href {https://doi.org/10.1103/PhysRevLett.134.116603} {\bibfield  {journal} {\bibinfo  {journal} {Phys. Rev. Lett.}\ }\textbf {\bibinfo {volume} {134}},\ \bibinfo {pages} {116603} (\bibinfo {year} {2025})}\BibitemShut {NoStop}%
\bibitem [{\citenamefont {Li}\ \emph {et~al.}(2021)\citenamefont {Li}, \citenamefont {Chen}, \citenamefont {Jiang},\ and\ \citenamefont {Xie}}]{PhysRevLett.127.236402}%
  \BibitemOpen
  \bibfield  {author} {\bibinfo {author} {\bibfnamefont {H.}~\bibnamefont {Li}}, \bibinfo {author} {\bibfnamefont {C.-Z.}\ \bibnamefont {Chen}}, \bibinfo {author} {\bibfnamefont {H.}~\bibnamefont {Jiang}},\ and\ \bibinfo {author} {\bibfnamefont {X.~C.}\ \bibnamefont {Xie}},\ }\bibfield  {title} {\bibinfo {title} {Coexistence of quantum hall and quantum anomalous hall phases in disordered {MnBi\textsubscript{2}Te\textsubscript{4}}},\ }\href {https://doi.org/10.1103/PhysRevLett.127.236402} {\bibfield  {journal} {\bibinfo  {journal} {Phys. Rev. Lett.}\ }\textbf {\bibinfo {volume} {127}},\ \bibinfo {pages} {236402} (\bibinfo {year} {2021})}\BibitemShut {NoStop}%
\bibitem [{\citenamefont {Li}\ \emph {et~al.}(2023)\citenamefont {Li}, \citenamefont {Liu}, \citenamefont {Liu}, \citenamefont {Wang}, \citenamefont {Lu},\ and\ \citenamefont {Xie}}]{10.1093/nsr/nwac296}%
  \BibitemOpen
  \bibfield  {author} {\bibinfo {author} {\bibfnamefont {S.}~\bibnamefont {Li}}, \bibinfo {author} {\bibfnamefont {T.}~\bibnamefont {Liu}}, \bibinfo {author} {\bibfnamefont {C.}~\bibnamefont {Liu}}, \bibinfo {author} {\bibfnamefont {Y.}~\bibnamefont {Wang}}, \bibinfo {author} {\bibfnamefont {H.-Z.}\ \bibnamefont {Lu}},\ and\ \bibinfo {author} {\bibfnamefont {X.~C.}\ \bibnamefont {Xie}},\ }\bibfield  {title} {\bibinfo {title} {Progress on the antiferromagnetic topological insulator {MnBi\textsubscript{2}Te\textsubscript{4}}},\ }\href {https://doi.org/10.1093/nsr/nwac296} {\bibfield  {journal} {\bibinfo  {journal} {National Science Review}\ }\textbf {\bibinfo {volume} {11}},\ \bibinfo {pages} {nwac296} (\bibinfo {year} {2023})}\BibitemShut {NoStop}%
\bibitem [{\citenamefont {Cai}\ \emph {et~al.}(2022)\citenamefont {Cai}, \citenamefont {Ovchinnikov}, \citenamefont {Fei}, \citenamefont {He}, \citenamefont {Song}, \citenamefont {Lin}, \citenamefont {Wang}, \citenamefont {Cobden}, \citenamefont {Chu}, \citenamefont {Cui}, \citenamefont {Chang}, \citenamefont {Xiao}, \citenamefont {Yan},\ and\ \citenamefont {Xu}}]{cai_electric_2022}%
  \BibitemOpen
  \bibfield  {author} {\bibinfo {author} {\bibfnamefont {J.}~\bibnamefont {Cai}}, \bibinfo {author} {\bibfnamefont {D.}~\bibnamefont {Ovchinnikov}}, \bibinfo {author} {\bibfnamefont {Z.}~\bibnamefont {Fei}}, \bibinfo {author} {\bibfnamefont {M.}~\bibnamefont {He}}, \bibinfo {author} {\bibfnamefont {T.}~\bibnamefont {Song}}, \bibinfo {author} {\bibfnamefont {Z.}~\bibnamefont {Lin}}, \bibinfo {author} {\bibfnamefont {C.}~\bibnamefont {Wang}}, \bibinfo {author} {\bibfnamefont {D.}~\bibnamefont {Cobden}}, \bibinfo {author} {\bibfnamefont {J.-H.}\ \bibnamefont {Chu}}, \bibinfo {author} {\bibfnamefont {Y.-T.}\ \bibnamefont {Cui}}, \bibinfo {author} {\bibfnamefont {C.-Z.}\ \bibnamefont {Chang}}, \bibinfo {author} {\bibfnamefont {D.}~\bibnamefont {Xiao}}, \bibinfo {author} {\bibfnamefont {J.}~\bibnamefont {Yan}},\ and\ \bibinfo {author} {\bibfnamefont {X.}~\bibnamefont {Xu}},\ }\bibfield  {title} {\bibinfo {title} {Electric control of a canted-antiferromagnetic {Chern} insulator},\ }\href
  {https://doi.org/10.1038/s41467-022-29259-8} {\bibfield  {journal} {\bibinfo  {journal} {Nature Communications}\ }\textbf {\bibinfo {volume} {13}},\ \bibinfo {pages} {1668} (\bibinfo {year} {2022})}\BibitemShut {NoStop}%
\bibitem [{\citenamefont {Sun}\ \emph {et~al.}(2020)\citenamefont {Sun}, \citenamefont {Wang}, \citenamefont {Zhang}, \citenamefont {Chen}, \citenamefont {Zhao}, \citenamefont {Liu}, \citenamefont {Liu}, \citenamefont {Chen}, \citenamefont {Lu},\ and\ \citenamefont {Xie}}]{PhysRevB.102.241406}%
  \BibitemOpen
  \bibfield  {author} {\bibinfo {author} {\bibfnamefont {H.-P.}\ \bibnamefont {Sun}}, \bibinfo {author} {\bibfnamefont {C.~M.}\ \bibnamefont {Wang}}, \bibinfo {author} {\bibfnamefont {S.-B.}\ \bibnamefont {Zhang}}, \bibinfo {author} {\bibfnamefont {R.}~\bibnamefont {Chen}}, \bibinfo {author} {\bibfnamefont {Y.}~\bibnamefont {Zhao}}, \bibinfo {author} {\bibfnamefont {C.}~\bibnamefont {Liu}}, \bibinfo {author} {\bibfnamefont {Q.}~\bibnamefont {Liu}}, \bibinfo {author} {\bibfnamefont {C.}~\bibnamefont {Chen}}, \bibinfo {author} {\bibfnamefont {H.-Z.}\ \bibnamefont {Lu}},\ and\ \bibinfo {author} {\bibfnamefont {X.~C.}\ \bibnamefont {Xie}},\ }\bibfield  {title} {\bibinfo {title} {Analytical solution for the surface states of the antiferromagnetic topological insulator {MnBi\textsubscript{2}Te\textsubscript{4}}},\ }\href {https://doi.org/10.1103/PhysRevB.102.241406} {\bibfield  {journal} {\bibinfo  {journal} {Phys. Rev. B}\ }\textbf {\bibinfo {volume} {102}},\ \bibinfo {pages} {241406} (\bibinfo {year}
  {2020})}\BibitemShut {NoStop}%
\bibitem [{\citenamefont {Ge}\ \emph {et~al.}(2020)\citenamefont {Ge}, \citenamefont {Liu}, \citenamefont {Li}, \citenamefont {Li}, \citenamefont {Luo}, \citenamefont {Wu}, \citenamefont {Xu},\ and\ \citenamefont {Wang}}]{10.1093/nsr/nwaa089}%
  \BibitemOpen
  \bibfield  {author} {\bibinfo {author} {\bibfnamefont {J.}~\bibnamefont {Ge}}, \bibinfo {author} {\bibfnamefont {Y.}~\bibnamefont {Liu}}, \bibinfo {author} {\bibfnamefont {J.}~\bibnamefont {Li}}, \bibinfo {author} {\bibfnamefont {H.}~\bibnamefont {Li}}, \bibinfo {author} {\bibfnamefont {T.}~\bibnamefont {Luo}}, \bibinfo {author} {\bibfnamefont {Y.}~\bibnamefont {Wu}}, \bibinfo {author} {\bibfnamefont {Y.}~\bibnamefont {Xu}},\ and\ \bibinfo {author} {\bibfnamefont {J.}~\bibnamefont {Wang}},\ }\bibfield  {title} {\bibinfo {title} {High-chern-number and high-temperature quantum hall effect without landau levels},\ }\href {https://doi.org/10.1093/nsr/nwaa089} {\bibfield  {journal} {\bibinfo  {journal} {National Science Review}\ }\textbf {\bibinfo {volume} {7}},\ \bibinfo {pages} {1280} (\bibinfo {year} {2020})}\BibitemShut {NoStop}%
\bibitem [{\citenamefont {Xu}\ \emph {et~al.}(2022)\citenamefont {Xu}, \citenamefont {Bai}, \citenamefont {Zhou}, \citenamefont {Li}, \citenamefont {Gu}, \citenamefont {Qin}, \citenamefont {Yin}, \citenamefont {Du}, \citenamefont {Zhang}, \citenamefont {Zhao}, \citenamefont {Li}, \citenamefont {Wu}, \citenamefont {Ding}, \citenamefont {Wang}, \citenamefont {Liang}, \citenamefont {Liu}, \citenamefont {Xu}, \citenamefont {Feng}, \citenamefont {He}, \citenamefont {Chen},\ and\ \citenamefont {Yang}}]{doi:10.1021/acs.nanolett.2c02034}%
  \BibitemOpen
  \bibfield  {author} {\bibinfo {author} {\bibfnamefont {R.}~\bibnamefont {Xu}}, \bibinfo {author} {\bibfnamefont {Y.}~\bibnamefont {Bai}}, \bibinfo {author} {\bibfnamefont {J.}~\bibnamefont {Zhou}}, \bibinfo {author} {\bibfnamefont {J.}~\bibnamefont {Li}}, \bibinfo {author} {\bibfnamefont {X.}~\bibnamefont {Gu}}, \bibinfo {author} {\bibfnamefont {N.}~\bibnamefont {Qin}}, \bibinfo {author} {\bibfnamefont {Z.}~\bibnamefont {Yin}}, \bibinfo {author} {\bibfnamefont {X.}~\bibnamefont {Du}}, \bibinfo {author} {\bibfnamefont {Q.}~\bibnamefont {Zhang}}, \bibinfo {author} {\bibfnamefont {W.}~\bibnamefont {Zhao}}, \bibinfo {author} {\bibfnamefont {Y.}~\bibnamefont {Li}}, \bibinfo {author} {\bibfnamefont {Y.}~\bibnamefont {Wu}}, \bibinfo {author} {\bibfnamefont {C.}~\bibnamefont {Ding}}, \bibinfo {author} {\bibfnamefont {L.}~\bibnamefont {Wang}}, \bibinfo {author} {\bibfnamefont {A.}~\bibnamefont {Liang}}, \bibinfo {author} {\bibfnamefont {Z.}~\bibnamefont {Liu}}, \bibinfo {author} {\bibfnamefont {Y.}~\bibnamefont
  {Xu}}, \bibinfo {author} {\bibfnamefont {X.}~\bibnamefont {Feng}}, \bibinfo {author} {\bibfnamefont {K.}~\bibnamefont {He}}, \bibinfo {author} {\bibfnamefont {Y.}~\bibnamefont {Chen}},\ and\ \bibinfo {author} {\bibfnamefont {L.}~\bibnamefont {Yang}},\ }\bibfield  {title} {\bibinfo {title} {Evolution of the electronic structure of ultrathin {MnBi\textsubscript{2}Te\textsubscript{4}} films},\ }\href {https://doi.org/10.1021/acs.nanolett.2c02034} {\bibfield  {journal} {\bibinfo  {journal} {Nano Letters}\ }\textbf {\bibinfo {volume} {22}},\ \bibinfo {pages} {6320} (\bibinfo {year} {2022})}\BibitemShut {NoStop}%
\bibitem [{\citenamefont {Lei}\ \emph {et~al.}(2021)\citenamefont {Lei}, \citenamefont {Heinonen}, \citenamefont {MacDonald},\ and\ \citenamefont {McQueeney}}]{PhysRevMaterials.5.064201}%
  \BibitemOpen
  \bibfield  {author} {\bibinfo {author} {\bibfnamefont {C.}~\bibnamefont {Lei}}, \bibinfo {author} {\bibfnamefont {O.}~\bibnamefont {Heinonen}}, \bibinfo {author} {\bibfnamefont {A.~H.}\ \bibnamefont {MacDonald}},\ and\ \bibinfo {author} {\bibfnamefont {R.~J.}\ \bibnamefont {McQueeney}},\ }\bibfield  {title} {\bibinfo {title} {Metamagnetism of few-layer topological antiferromagnets},\ }\href {https://doi.org/10.1103/PhysRevMaterials.5.064201} {\bibfield  {journal} {\bibinfo  {journal} {Phys. Rev. Mater.}\ }\textbf {\bibinfo {volume} {5}},\ \bibinfo {pages} {064201} (\bibinfo {year} {2021})}\BibitemShut {NoStop}%
\bibitem [{\citenamefont {Ghosh}\ \emph {et~al.}(2024)\citenamefont {Ghosh}, \citenamefont {Onishi}, \citenamefont {Xu}, \citenamefont {Lin}, \citenamefont {Fu},\ and\ \citenamefont {Bansil}}]{doi:10.1126/sciadv.ado1761}%
  \BibitemOpen
  \bibfield  {author} {\bibinfo {author} {\bibfnamefont {B.}~\bibnamefont {Ghosh}}, \bibinfo {author} {\bibfnamefont {Y.}~\bibnamefont {Onishi}}, \bibinfo {author} {\bibfnamefont {S.-Y.}\ \bibnamefont {Xu}}, \bibinfo {author} {\bibfnamefont {H.}~\bibnamefont {Lin}}, \bibinfo {author} {\bibfnamefont {L.}~\bibnamefont {Fu}},\ and\ \bibinfo {author} {\bibfnamefont {A.}~\bibnamefont {Bansil}},\ }\bibfield  {title} {\bibinfo {title} {Probing quantum geometry through optical conductivity and magnetic circular dichroism},\ }\href {https://doi.org/10.1126/sciadv.ado1761} {\bibfield  {journal} {\bibinfo  {journal} {Science Advances}\ }\textbf {\bibinfo {volume} {10}},\ \bibinfo {pages} {eado1761} (\bibinfo {year} {2024})}\BibitemShut {NoStop}%
\bibitem [{\citenamefont {Kattan}\ \emph {et~al.}(2025)\citenamefont {Kattan}, \citenamefont {Duff},\ and\ \citenamefont {Sipe}}]{PhysRevB.111.075202}%
  \BibitemOpen
  \bibfield  {author} {\bibinfo {author} {\bibfnamefont {J.~G.}\ \bibnamefont {Kattan}}, \bibinfo {author} {\bibfnamefont {A.~H.}\ \bibnamefont {Duff}},\ and\ \bibinfo {author} {\bibfnamefont {J.~E.}\ \bibnamefont {Sipe}},\ }\bibfield  {title} {\bibinfo {title} {Linear response of a {Chern} insulator to finite-frequency electric fields},\ }\href {https://doi.org/10.1103/PhysRevB.111.075202} {\bibfield  {journal} {\bibinfo  {journal} {Phys. Rev. B}\ }\textbf {\bibinfo {volume} {111}},\ \bibinfo {pages} {075202} (\bibinfo {year} {2025})}\BibitemShut {NoStop}%
\bibitem [{\citenamefont {Mahon}\ \emph {et~al.}(2023)\citenamefont {Mahon}, \citenamefont {Kattan},\ and\ \citenamefont {Sipe}}]{mahon2023polarization}%
  \BibitemOpen
  \bibfield  {author} {\bibinfo {author} {\bibfnamefont {P.~T.}\ \bibnamefont {Mahon}}, \bibinfo {author} {\bibfnamefont {J.~G.}\ \bibnamefont {Kattan}},\ and\ \bibinfo {author} {\bibfnamefont {J.~E.}\ \bibnamefont {Sipe}},\ }\bibfield  {title} {\bibinfo {title} {Polarization and orbital magnetization in {Chern} insulators: A microscopic perspective},\ }\href@noop {} {\bibfield  {journal} {\bibinfo  {journal} {Physical Review B}\ }\textbf {\bibinfo {volume} {107}},\ \bibinfo {pages} {115110} (\bibinfo {year} {2023})}\BibitemShut {NoStop}%
\bibitem [{\citenamefont {Kattan}\ and\ \citenamefont {Sipe}(2026)}]{kattan2025chern}%
  \BibitemOpen
  \bibfield  {author} {\bibinfo {author} {\bibfnamefont {J.~G.}\ \bibnamefont {Kattan}}\ and\ \bibinfo {author} {\bibfnamefont {J.~E.}\ \bibnamefont {Sipe}},\ }\bibfield  {title} {\bibinfo {title} {Chern insulators in two and three dimensions: A global perspective},\ }\href {https://doi.org/10.1103/4dbx-5bxj} {\bibfield  {journal} {\bibinfo  {journal} {Phys. Rev. B}\ }\textbf {\bibinfo {volume} {113}},\ \bibinfo {pages} {125201} (\bibinfo {year} {2026})}\BibitemShut {NoStop}%
\bibitem [{\citenamefont {Mahon}\ \emph {et~al.}(2019)\citenamefont {Mahon}, \citenamefont {Muniz},\ and\ \citenamefont {Sipe}}]{PhysRevB.99.235140}%
  \BibitemOpen
  \bibfield  {author} {\bibinfo {author} {\bibfnamefont {P.~T.}\ \bibnamefont {Mahon}}, \bibinfo {author} {\bibfnamefont {R.~A.}\ \bibnamefont {Muniz}},\ and\ \bibinfo {author} {\bibfnamefont {J.~E.}\ \bibnamefont {Sipe}},\ }\bibfield  {title} {\bibinfo {title} {Microscopic polarization and magnetization fields in extended systems},\ }\href {https://doi.org/10.1103/PhysRevB.99.235140} {\bibfield  {journal} {\bibinfo  {journal} {Phys. Rev. B}\ }\textbf {\bibinfo {volume} {99}},\ \bibinfo {pages} {235140} (\bibinfo {year} {2019})}\BibitemShut {NoStop}%
\bibitem [{\citenamefont {Mu\~noz Segovia}\ \emph {et~al.}(2025)\citenamefont {Mu\~noz Segovia}, \citenamefont {Cr\'epel}, \citenamefont {Queiroz},\ and\ \citenamefont {Millis}}]{m6c2-yd5j}%
  \BibitemOpen
  \bibfield  {author} {\bibinfo {author} {\bibfnamefont {D.}~\bibnamefont {Mu\~noz Segovia}}, \bibinfo {author} {\bibfnamefont {V.}~\bibnamefont {Cr\'epel}}, \bibinfo {author} {\bibfnamefont {R.}~\bibnamefont {Queiroz}},\ and\ \bibinfo {author} {\bibfnamefont {A.~J.}\ \bibnamefont {Millis}},\ }\bibfield  {title} {\bibinfo {title} {Twist-angle evolution of the intervalley-coherent antiferromagnet in twisted ${\mathrm{wse}}_{2}$},\ }\href {https://doi.org/10.1103/m6c2-yd5j} {\bibfield  {journal} {\bibinfo  {journal} {Phys. Rev. B}\ }\textbf {\bibinfo {volume} {112}},\ \bibinfo {pages} {085111} (\bibinfo {year} {2025})}\BibitemShut {NoStop}%
\bibitem [{\citenamefont {Venderbos}\ \emph {et~al.}(2018)\citenamefont {Venderbos}, \citenamefont {Hu},\ and\ \citenamefont {Kane}}]{PhysRevB.98.235160}%
  \BibitemOpen
  \bibfield  {author} {\bibinfo {author} {\bibfnamefont {J.~W.~F.}\ \bibnamefont {Venderbos}}, \bibinfo {author} {\bibfnamefont {Y.}~\bibnamefont {Hu}},\ and\ \bibinfo {author} {\bibfnamefont {C.~L.}\ \bibnamefont {Kane}},\ }\bibfield  {title} {\bibinfo {title} {Higher angular momentum band inversions in two dimensions},\ }\href {https://doi.org/10.1103/PhysRevB.98.235160} {\bibfield  {journal} {\bibinfo  {journal} {Phys. Rev. B}\ }\textbf {\bibinfo {volume} {98}},\ \bibinfo {pages} {235160} (\bibinfo {year} {2018})}\BibitemShut {NoStop}%
\bibitem [{\citenamefont {Lopez}\ \emph {et~al.}(2025)\citenamefont {Lopez}, \citenamefont {Melton}, \citenamefont {Ahn}, \citenamefont {Rubenstein},\ and\ \citenamefont {Krogel}}]{doi:10.1021/acs.jctc.5c00838}%
  \BibitemOpen
  \bibfield  {author} {\bibinfo {author} {\bibfnamefont {A.}~\bibnamefont {Lopez}}, \bibinfo {author} {\bibfnamefont {C.~A.}\ \bibnamefont {Melton}}, \bibinfo {author} {\bibfnamefont {J.}~\bibnamefont {Ahn}}, \bibinfo {author} {\bibfnamefont {B.~M.}\ \bibnamefont {Rubenstein}},\ and\ \bibinfo {author} {\bibfnamefont {J.~T.}\ \bibnamefont {Krogel}},\ }\bibfield  {title} {\bibinfo {title} {Identifying band inversions in topological materials using diffusion monte carlo},\ }\href {https://doi.org/10.1021/acs.jctc.5c00838} {\bibfield  {journal} {\bibinfo  {journal} {Journal of Chemical Theory and Computation}\ }\textbf {\bibinfo {volume} {21}},\ \bibinfo {pages} {7485} (\bibinfo {year} {2025})}\BibitemShut {NoStop}%
\bibitem [{\citenamefont {Bl\"ochl}(1994)}]{PhysRevB.50.17953}%
  \BibitemOpen
  \bibfield  {author} {\bibinfo {author} {\bibfnamefont {P.~E.}\ \bibnamefont {Bl\"ochl}},\ }\bibfield  {title} {\bibinfo {title} {Projector augmented-wave method},\ }\href {https://doi.org/10.1103/PhysRevB.50.17953} {\bibfield  {journal} {\bibinfo  {journal} {Phys. Rev. B}\ }\textbf {\bibinfo {volume} {50}},\ \bibinfo {pages} {17953} (\bibinfo {year} {1994})}\BibitemShut {NoStop}%
\bibitem [{\citenamefont {Kresse}\ and\ \citenamefont {Joubert}(1999)}]{PhysRevB.59.1758}%
  \BibitemOpen
  \bibfield  {author} {\bibinfo {author} {\bibfnamefont {G.}~\bibnamefont {Kresse}}\ and\ \bibinfo {author} {\bibfnamefont {D.}~\bibnamefont {Joubert}},\ }\bibfield  {title} {\bibinfo {title} {From ultrasoft pseudopotentials to the projector augmented-wave method},\ }\href {https://doi.org/10.1103/PhysRevB.59.1758} {\bibfield  {journal} {\bibinfo  {journal} {Phys. Rev. B}\ }\textbf {\bibinfo {volume} {59}},\ \bibinfo {pages} {1758} (\bibinfo {year} {1999})}\BibitemShut {NoStop}%
\bibitem [{\citenamefont {Kresse}\ and\ \citenamefont {Hafner}(1993)}]{PhysRevB.47.558}%
  \BibitemOpen
  \bibfield  {author} {\bibinfo {author} {\bibfnamefont {G.}~\bibnamefont {Kresse}}\ and\ \bibinfo {author} {\bibfnamefont {J.}~\bibnamefont {Hafner}},\ }\bibfield  {title} {\bibinfo {title} {Ab initio molecular dynamics for liquid metals},\ }\href {https://doi.org/10.1103/PhysRevB.47.558} {\bibfield  {journal} {\bibinfo  {journal} {Phys. Rev. B}\ }\textbf {\bibinfo {volume} {47}},\ \bibinfo {pages} {558} (\bibinfo {year} {1993})}\BibitemShut {NoStop}%
\bibitem [{\citenamefont {Kresse}\ and\ \citenamefont {Furthm\"uller}(1996)}]{PhysRevB.54.11169}%
  \BibitemOpen
  \bibfield  {author} {\bibinfo {author} {\bibfnamefont {G.}~\bibnamefont {Kresse}}\ and\ \bibinfo {author} {\bibfnamefont {J.}~\bibnamefont {Furthm\"uller}},\ }\bibfield  {title} {\bibinfo {title} {Efficient iterative schemes for ab initio total-energy calculations using a plane-wave basis set},\ }\href {https://doi.org/10.1103/PhysRevB.54.11169} {\bibfield  {journal} {\bibinfo  {journal} {Phys. Rev. B}\ }\textbf {\bibinfo {volume} {54}},\ \bibinfo {pages} {11169} (\bibinfo {year} {1996})}\BibitemShut {NoStop}%
\bibitem [{\citenamefont {Kresse}\ and\ \citenamefont {Furthmüller}(1996)}]{KRESSE199615}%
  \BibitemOpen
  \bibfield  {author} {\bibinfo {author} {\bibfnamefont {G.}~\bibnamefont {Kresse}}\ and\ \bibinfo {author} {\bibfnamefont {J.}~\bibnamefont {Furthmüller}},\ }\bibfield  {title} {\bibinfo {title} {Efficiency of ab-initio total energy calculations for metals and semiconductors using a plane-wave basis set},\ }\href {https://doi.org/https://doi.org/10.1016/0927-0256(96)00008-0} {\bibfield  {journal} {\bibinfo  {journal} {Comput. Mater. Sci.}\ }\textbf {\bibinfo {volume} {6}},\ \bibinfo {pages} {15} (\bibinfo {year} {1996})}\BibitemShut {NoStop}%
\bibitem [{\citenamefont {Perdew}\ \emph {et~al.}(1996)\citenamefont {Perdew}, \citenamefont {Burke},\ and\ \citenamefont {Ernzerhof}}]{PhysRevLett.77.3865}%
  \BibitemOpen
  \bibfield  {author} {\bibinfo {author} {\bibfnamefont {J.~P.}\ \bibnamefont {Perdew}}, \bibinfo {author} {\bibfnamefont {K.}~\bibnamefont {Burke}},\ and\ \bibinfo {author} {\bibfnamefont {M.}~\bibnamefont {Ernzerhof}},\ }\bibfield  {title} {\bibinfo {title} {Generalized gradient approximation made simple},\ }\href {https://doi.org/10.1103/PhysRevLett.77.3865} {\bibfield  {journal} {\bibinfo  {journal} {Phys. Rev. Lett.}\ }\textbf {\bibinfo {volume} {77}},\ \bibinfo {pages} {3865} (\bibinfo {year} {1996})}\BibitemShut {NoStop}%
\bibitem [{\citenamefont {Liu}\ \emph {et~al.}(2022)\citenamefont {Liu}, \citenamefont {Lei}, \citenamefont {Kim}, \citenamefont {Li}, \citenamefont {Frammolino}, \citenamefont {Yan}, \citenamefont {Macdonald},\ and\ \citenamefont {Shih}}]{doi:10.1073/pnas.2207681119}%
  \BibitemOpen
  \bibfield  {author} {\bibinfo {author} {\bibfnamefont {M.}~\bibnamefont {Liu}}, \bibinfo {author} {\bibfnamefont {C.}~\bibnamefont {Lei}}, \bibinfo {author} {\bibfnamefont {H.}~\bibnamefont {Kim}}, \bibinfo {author} {\bibfnamefont {Y.}~\bibnamefont {Li}}, \bibinfo {author} {\bibfnamefont {L.}~\bibnamefont {Frammolino}}, \bibinfo {author} {\bibfnamefont {J.}~\bibnamefont {Yan}}, \bibinfo {author} {\bibfnamefont {A.~H.}\ \bibnamefont {Macdonald}},\ and\ \bibinfo {author} {\bibfnamefont {C.-K.}\ \bibnamefont {Shih}},\ }\bibfield  {title} {\bibinfo {title} {Visualizing the interplay of dirac mass gap and magnetism at nanoscale in intrinsic magnetic topological insulators},\ }\href {https://doi.org/10.1073/pnas.2207681119} {\bibfield  {journal} {\bibinfo  {journal} {Proceedings of the National Academy of Sciences}\ }\textbf {\bibinfo {volume} {119}},\ \bibinfo {pages} {e2207681119} (\bibinfo {year} {2022})}\BibitemShut {NoStop}%
\bibitem [{\citenamefont {Dudarev}\ \emph {et~al.}(1998)\citenamefont {Dudarev}, \citenamefont {Botton}, \citenamefont {Savrasov}, \citenamefont {Humphreys},\ and\ \citenamefont {Sutton}}]{PhysRevB.57.1505}%
  \BibitemOpen
  \bibfield  {author} {\bibinfo {author} {\bibfnamefont {S.~L.}\ \bibnamefont {Dudarev}}, \bibinfo {author} {\bibfnamefont {G.~A.}\ \bibnamefont {Botton}}, \bibinfo {author} {\bibfnamefont {S.~Y.}\ \bibnamefont {Savrasov}}, \bibinfo {author} {\bibfnamefont {C.~J.}\ \bibnamefont {Humphreys}},\ and\ \bibinfo {author} {\bibfnamefont {A.~P.}\ \bibnamefont {Sutton}},\ }\bibfield  {title} {\bibinfo {title} {Electron-energy-loss spectra and the structural stability of nickel oxide: An lsda+u study},\ }\href {https://doi.org/10.1103/PhysRevB.57.1505} {\bibfield  {journal} {\bibinfo  {journal} {Phys. Rev. B}\ }\textbf {\bibinfo {volume} {57}},\ \bibinfo {pages} {1505} (\bibinfo {year} {1998})}\BibitemShut {NoStop}%
\bibitem [{\citenamefont {Zeugner}\ \emph {et~al.}(2019)\citenamefont {Zeugner}, \citenamefont {Nietschke}, \citenamefont {Wolter}, \citenamefont {Gaß}, \citenamefont {Vidal}, \citenamefont {Peixoto}, \citenamefont {Pohl}, \citenamefont {Damm}, \citenamefont {Lubk}, \citenamefont {Hentrich}, \citenamefont {Moser}, \citenamefont {Fornari}, \citenamefont {Min}, \citenamefont {Schatz}, \citenamefont {Kißner}, \citenamefont {{\"U}nzelmann}, \citenamefont {Kaiser}, \citenamefont {Scaravaggi}, \citenamefont {Rellinghaus}, \citenamefont {Nielsch}, \citenamefont {Hess}, \citenamefont {B{\"u}chner}, \citenamefont {Reinert}, \citenamefont {Bentmann}, \citenamefont {Oeckler}, \citenamefont {Doert}, \citenamefont {Ruck},\ and\ \citenamefont {Isaeva}}]{doi:10.1021/acs.chemmater.8b05017}%
  \BibitemOpen
  \bibfield  {author} {\bibinfo {author} {\bibfnamefont {A.}~\bibnamefont {Zeugner}}, \bibinfo {author} {\bibfnamefont {F.}~\bibnamefont {Nietschke}}, \bibinfo {author} {\bibfnamefont {A.~U.~B.}\ \bibnamefont {Wolter}}, \bibinfo {author} {\bibfnamefont {S.}~\bibnamefont {Gaß}}, \bibinfo {author} {\bibfnamefont {R.~C.}\ \bibnamefont {Vidal}}, \bibinfo {author} {\bibfnamefont {T.~R.~F.}\ \bibnamefont {Peixoto}}, \bibinfo {author} {\bibfnamefont {D.}~\bibnamefont {Pohl}}, \bibinfo {author} {\bibfnamefont {C.}~\bibnamefont {Damm}}, \bibinfo {author} {\bibfnamefont {A.}~\bibnamefont {Lubk}}, \bibinfo {author} {\bibfnamefont {R.}~\bibnamefont {Hentrich}}, \bibinfo {author} {\bibfnamefont {S.~K.}\ \bibnamefont {Moser}}, \bibinfo {author} {\bibfnamefont {C.}~\bibnamefont {Fornari}}, \bibinfo {author} {\bibfnamefont {C.~H.}\ \bibnamefont {Min}}, \bibinfo {author} {\bibfnamefont {S.}~\bibnamefont {Schatz}}, \bibinfo {author} {\bibfnamefont {K.}~\bibnamefont {Kißner}}, \bibinfo {author} {\bibfnamefont
  {M.}~\bibnamefont {{\"U}nzelmann}}, \bibinfo {author} {\bibfnamefont {M.}~\bibnamefont {Kaiser}}, \bibinfo {author} {\bibfnamefont {F.}~\bibnamefont {Scaravaggi}}, \bibinfo {author} {\bibfnamefont {B.}~\bibnamefont {Rellinghaus}}, \bibinfo {author} {\bibfnamefont {K.}~\bibnamefont {Nielsch}}, \bibinfo {author} {\bibfnamefont {C.}~\bibnamefont {Hess}}, \bibinfo {author} {\bibfnamefont {B.}~\bibnamefont {B{\"u}chner}}, \bibinfo {author} {\bibfnamefont {F.}~\bibnamefont {Reinert}}, \bibinfo {author} {\bibfnamefont {H.}~\bibnamefont {Bentmann}}, \bibinfo {author} {\bibfnamefont {O.}~\bibnamefont {Oeckler}}, \bibinfo {author} {\bibfnamefont {T.}~\bibnamefont {Doert}}, \bibinfo {author} {\bibfnamefont {M.}~\bibnamefont {Ruck}},\ and\ \bibinfo {author} {\bibfnamefont {A.}~\bibnamefont {Isaeva}},\ }\bibfield  {title} {\bibinfo {title} {Chemical aspects of the candidate antiferromagnetic topological insulator {MnBi\textsubscript{2}Te\textsubscript{4}}},\ }\href {https://doi.org/10.1021/acs.chemmater.8b05017}
  {\bibfield  {journal} {\bibinfo  {journal} {Chemistry of Materials}\ }\textbf {\bibinfo {volume} {31}},\ \bibinfo {pages} {2795} (\bibinfo {year} {2019})},\ \Eprint {https://arxiv.org/abs/https://doi.org/10.1021/acs.chemmater.8b05017} {https://doi.org/10.1021/acs.chemmater.8b05017} \BibitemShut {NoStop}%
\bibitem [{\citenamefont {Bloch}(1929)}]{bloch_uber_1929}%
  \BibitemOpen
  \bibfield  {author} {\bibinfo {author} {\bibfnamefont {F.}~\bibnamefont {Bloch}},\ }\bibfield  {title} {\bibinfo {title} {Über die {Quantenmechanik} der {Elektronen} in {Kristallgittern}},\ }\href {https://doi.org/10.1007/BF01339455} {\bibfield  {journal} {\bibinfo  {journal} {Zeitschrift für Physik}\ }\textbf {\bibinfo {volume} {52}},\ \bibinfo {pages} {555} (\bibinfo {year} {1929})}\BibitemShut {NoStop}%
\bibitem [{\citenamefont {Marzari}\ \emph {et~al.}(2012)\citenamefont {Marzari}, \citenamefont {Mostofi}, \citenamefont {Yates}, \citenamefont {Souza},\ and\ \citenamefont {Vanderbilt}}]{RevModPhys.84.1419}%
  \BibitemOpen
  \bibfield  {author} {\bibinfo {author} {\bibfnamefont {N.}~\bibnamefont {Marzari}}, \bibinfo {author} {\bibfnamefont {A.~A.}\ \bibnamefont {Mostofi}}, \bibinfo {author} {\bibfnamefont {J.~R.}\ \bibnamefont {Yates}}, \bibinfo {author} {\bibfnamefont {I.}~\bibnamefont {Souza}},\ and\ \bibinfo {author} {\bibfnamefont {D.}~\bibnamefont {Vanderbilt}},\ }\bibfield  {title} {\bibinfo {title} {Maximally localized wannier functions: Theory and applications},\ }\href {https://doi.org/10.1103/RevModPhys.84.1419} {\bibfield  {journal} {\bibinfo  {journal} {Rev. Mod. Phys.}\ }\textbf {\bibinfo {volume} {84}},\ \bibinfo {pages} {1419} (\bibinfo {year} {2012})}\BibitemShut {NoStop}%
\bibitem [{\citenamefont {Wang}\ \emph {et~al.}(2006)\citenamefont {Wang}, \citenamefont {Yates}, \citenamefont {Souza},\ and\ \citenamefont {Vanderbilt}}]{PhysRevB.74.195118}%
  \BibitemOpen
  \bibfield  {author} {\bibinfo {author} {\bibfnamefont {X.}~\bibnamefont {Wang}}, \bibinfo {author} {\bibfnamefont {J.~R.}\ \bibnamefont {Yates}}, \bibinfo {author} {\bibfnamefont {I.}~\bibnamefont {Souza}},\ and\ \bibinfo {author} {\bibfnamefont {D.}~\bibnamefont {Vanderbilt}},\ }\bibfield  {title} {\bibinfo {title} {Ab initio calculation of the anomalous hall conductivity by wannier interpolation},\ }\href {https://doi.org/10.1103/PhysRevB.74.195118} {\bibfield  {journal} {\bibinfo  {journal} {Phys. Rev. B}\ }\textbf {\bibinfo {volume} {74}},\ \bibinfo {pages} {195118} (\bibinfo {year} {2006})}\BibitemShut {NoStop}%
\bibitem [{\citenamefont {Yates}\ \emph {et~al.}(2007)\citenamefont {Yates}, \citenamefont {Wang}, \citenamefont {Vanderbilt},\ and\ \citenamefont {Souza}}]{PhysRevB.75.195121}%
  \BibitemOpen
  \bibfield  {author} {\bibinfo {author} {\bibfnamefont {J.~R.}\ \bibnamefont {Yates}}, \bibinfo {author} {\bibfnamefont {X.}~\bibnamefont {Wang}}, \bibinfo {author} {\bibfnamefont {D.}~\bibnamefont {Vanderbilt}},\ and\ \bibinfo {author} {\bibfnamefont {I.}~\bibnamefont {Souza}},\ }\bibfield  {title} {\bibinfo {title} {Spectral and fermi surface properties from wannier interpolation},\ }\href {https://doi.org/10.1103/PhysRevB.75.195121} {\bibfield  {journal} {\bibinfo  {journal} {Phys. Rev. B}\ }\textbf {\bibinfo {volume} {75}},\ \bibinfo {pages} {195121} (\bibinfo {year} {2007})}\BibitemShut {NoStop}%
\bibitem [{Note1()}]{Note1}%
  \BibitemOpen
  \bibinfo {note} {\label {fn:berry_connection}In the Wannier interpolation literature, the Berry connection in the Bloch basis is most commonly written as \begin {equation} \xi _{\ell , nm} (\protect \boldsymbol {k})=U_{n\alpha }(\protect \boldsymbol {k})\protect \tilde {\xi }_{\ell , \alpha \beta }(\protect \boldsymbol {k}) U^{\dagger }_{\beta m}(\protect \boldsymbol {k}) + i U_{n\alpha }(\protect \boldsymbol {k}) \partial _{\ell } U^{\dagger }_{\alpha m}(\protect \boldsymbol {k}). \label {eq:BlochBerry} \end {equation} The second term in Eq.~(\ref {eq:BlochBerry}) can be evaluated using $\protect \boldsymbol {k}\cdot \protect \boldsymbol {p}$ perturbation theory and is related to the anti-Hermitian matrix \begin {equation} (U(\protect \boldsymbol {k}) \partial _{\ell } U^{\dagger }(\protect \boldsymbol {k}))_{nm} = \begin {cases} \protect \dfrac {\protect \bar {H}_{\ell , nm}(\protect \boldsymbol {k})}{E_{m \protect \boldsymbol {k}} - E_{n \protect \boldsymbol {k}}}, & \protect \text {if } n \protect \neq
  m, \\[1em] 0, & \protect \text {if } n = m \label {eq:Ds} \end {cases} \end {equation} where we have defined $\protect \bar {H}_{\ell }(\protect \boldsymbol {k})=U (\protect \boldsymbol {k})\protect \tilde {H}_{\ell }(\protect \boldsymbol {k})U^\dagger (\protect \boldsymbol {k})$, etc. Note that Eq.~(\ref {eq:Ds}) diverges in the case of band crossings. However, here we only consider matrix elements between unoccupied and occupied states of an insulator.}\BibitemShut {Stop}%
\bibitem [{\citenamefont {Qiao}\ \emph {et~al.}(2018)\citenamefont {Qiao}, \citenamefont {Zhou}, \citenamefont {Yuan},\ and\ \citenamefont {Zhao}}]{PhysRevB.98.214402}%
  \BibitemOpen
  \bibfield  {author} {\bibinfo {author} {\bibfnamefont {J.}~\bibnamefont {Qiao}}, \bibinfo {author} {\bibfnamefont {J.}~\bibnamefont {Zhou}}, \bibinfo {author} {\bibfnamefont {Z.}~\bibnamefont {Yuan}},\ and\ \bibinfo {author} {\bibfnamefont {W.}~\bibnamefont {Zhao}},\ }\bibfield  {title} {\bibinfo {title} {Calculation of intrinsic spin hall conductivity by wannier interpolation},\ }\href {https://doi.org/10.1103/PhysRevB.98.214402} {\bibfield  {journal} {\bibinfo  {journal} {Phys. Rev. B}\ }\textbf {\bibinfo {volume} {98}},\ \bibinfo {pages} {214402} (\bibinfo {year} {2018})}\BibitemShut {NoStop}%
\bibitem [{\citenamefont {Li}\ \emph {et~al.}(2019)\citenamefont {Li}, \citenamefont {Li}, \citenamefont {Du}, \citenamefont {Wang}, \citenamefont {Gu}, \citenamefont {Zhang}, \citenamefont {He}, \citenamefont {Duan},\ and\ \citenamefont {Xu}}]{doi:10.1126/sciadv.aaw5685}%
  \BibitemOpen
  \bibfield  {author} {\bibinfo {author} {\bibfnamefont {J.}~\bibnamefont {Li}}, \bibinfo {author} {\bibfnamefont {Y.}~\bibnamefont {Li}}, \bibinfo {author} {\bibfnamefont {S.}~\bibnamefont {Du}}, \bibinfo {author} {\bibfnamefont {Z.}~\bibnamefont {Wang}}, \bibinfo {author} {\bibfnamefont {B.-L.}\ \bibnamefont {Gu}}, \bibinfo {author} {\bibfnamefont {S.-C.}\ \bibnamefont {Zhang}}, \bibinfo {author} {\bibfnamefont {K.}~\bibnamefont {He}}, \bibinfo {author} {\bibfnamefont {W.}~\bibnamefont {Duan}},\ and\ \bibinfo {author} {\bibfnamefont {Y.}~\bibnamefont {Xu}},\ }\bibfield  {title} {\bibinfo {title} {Intrinsic magnetic topological insulators in van der waals layered {MnBi\textsubscript{2}Te\textsubscript{4}}-family materials},\ }\href {https://doi.org/10.1126/sciadv.aaw5685} {\bibfield  {journal} {\bibinfo  {journal} {Science Advances}\ }\textbf {\bibinfo {volume} {5}},\ \bibinfo {pages} {eaaw5685} (\bibinfo {year} {2019})}\BibitemShut {NoStop}%
\bibitem [{\citenamefont {Zhang}\ \emph {et~al.}(2009)\citenamefont {Zhang}, \citenamefont {Liu}, \citenamefont {Qi}, \citenamefont {Dai}, \citenamefont {Fang},\ and\ \citenamefont {Zhang}}]{zhang_topological_2009}%
  \BibitemOpen
  \bibfield  {author} {\bibinfo {author} {\bibfnamefont {H.}~\bibnamefont {Zhang}}, \bibinfo {author} {\bibfnamefont {C.-X.}\ \bibnamefont {Liu}}, \bibinfo {author} {\bibfnamefont {X.-L.}\ \bibnamefont {Qi}}, \bibinfo {author} {\bibfnamefont {X.}~\bibnamefont {Dai}}, \bibinfo {author} {\bibfnamefont {Z.}~\bibnamefont {Fang}},\ and\ \bibinfo {author} {\bibfnamefont {S.-C.}\ \bibnamefont {Zhang}},\ }\bibfield  {title} {\bibinfo {title} {Topological insulators in {Bi2Se3}, {Bi2Te3} and {Sb2Te3} with a single {Dirac} cone on the surface},\ }\href {https://doi.org/10.1038/nphys1270} {\bibfield  {journal} {\bibinfo  {journal} {Nature Physics}\ }\textbf {\bibinfo {volume} {5}},\ \bibinfo {pages} {438} (\bibinfo {year} {2009})}\BibitemShut {NoStop}%
\bibitem [{Note2()}]{Note2}%
  \BibitemOpen
  \bibinfo {note} {Others have reported a Chern number of $+1$ in 5 SL MnBi$_2$Te$_4$ \cite {PhysRevResearch.7.023024,PhysRevB.108.125424}. This difference in sign arises from their choice of convention in the Kubo-Greenwood formula, rather than a ground-state calculation as in Eq.~(\ref {eq:ChernNumber})}\BibitemShut {NoStop}%
\bibitem [{\citenamefont {Po}\ \emph {et~al.}(2018)\citenamefont {Po}, \citenamefont {Watanabe},\ and\ \citenamefont {Vishwanath}}]{PhysRevLett.121.126402}%
  \BibitemOpen
  \bibfield  {author} {\bibinfo {author} {\bibfnamefont {H.~C.}\ \bibnamefont {Po}}, \bibinfo {author} {\bibfnamefont {H.}~\bibnamefont {Watanabe}},\ and\ \bibinfo {author} {\bibfnamefont {A.}~\bibnamefont {Vishwanath}},\ }\bibfield  {title} {\bibinfo {title} {Fragile topology and wannier obstructions},\ }\href {https://doi.org/10.1103/PhysRevLett.121.126402} {\bibfield  {journal} {\bibinfo  {journal} {Phys. Rev. Lett.}\ }\textbf {\bibinfo {volume} {121}},\ \bibinfo {pages} {126402} (\bibinfo {year} {2018})}\BibitemShut {NoStop}%
\bibitem [{\citenamefont {Mahon}\ and\ \citenamefont {Sipe}(2020)}]{PhysRevResearch.2.043110}%
  \BibitemOpen
  \bibfield  {author} {\bibinfo {author} {\bibfnamefont {P.~T.}\ \bibnamefont {Mahon}}\ and\ \bibinfo {author} {\bibfnamefont {J.~E.}\ \bibnamefont {Sipe}},\ }\bibfield  {title} {\bibinfo {title} {From magnetoelectric response to optical activity},\ }\href {https://doi.org/10.1103/PhysRevResearch.2.043110} {\bibfield  {journal} {\bibinfo  {journal} {Phys. Rev. Res.}\ }\textbf {\bibinfo {volume} {2}},\ \bibinfo {pages} {043110} (\bibinfo {year} {2020})}\BibitemShut {NoStop}%
\bibitem [{\citenamefont {Wan}\ \emph {et~al.}(2025)\citenamefont {Wan}, \citenamefont {Liu},\ and\ \citenamefont {Sun}}]{8vs2-jvc4}%
  \BibitemOpen
  \bibfield  {author} {\bibinfo {author} {\bibfnamefont {Y.-H.}\ \bibnamefont {Wan}}, \bibinfo {author} {\bibfnamefont {P.-Y.}\ \bibnamefont {Liu}},\ and\ \bibinfo {author} {\bibfnamefont {Q.-F.}\ \bibnamefont {Sun}},\ }\bibfield  {title} {\bibinfo {title} {Quantum anomalous hall effect in ferromagnetic metals},\ }\href {https://doi.org/10.1103/8vs2-jvc4} {\bibfield  {journal} {\bibinfo  {journal} {Phys. Rev. Lett.}\ }\textbf {\bibinfo {volume} {135}},\ \bibinfo {pages} {186302} (\bibinfo {year} {2025})}\BibitemShut {NoStop}%
\bibitem [{\citenamefont {Lifshitz}\ and\ \citenamefont {Pitaevskii}(1995)}]{lifshitz1995physical}%
  \BibitemOpen
  \bibfield  {author} {\bibinfo {author} {\bibfnamefont {E.}~\bibnamefont {Lifshitz}}\ and\ \bibinfo {author} {\bibfnamefont {L.}~\bibnamefont {Pitaevskii}},\ }\href {https://books.google.ca/books?id=HgRBAQAAIAAJ} {\emph {\bibinfo {title} {Physical Kinetics: Volume 10}}},\ Course of theoretical physics\ (\bibinfo  {publisher} {Elsevier Science},\ \bibinfo {year} {1995})\BibitemShut {NoStop}%
\bibitem [{\citenamefont {Lucarini}\ \emph {et~al.}(2005)\citenamefont {Lucarini}, \citenamefont {Peiponen}, \citenamefont {Saarinen},\ and\ \citenamefont {Vartiainen}}]{lucarini2005kramers}%
  \BibitemOpen
  \bibfield  {author} {\bibinfo {author} {\bibfnamefont {V.}~\bibnamefont {Lucarini}}, \bibinfo {author} {\bibfnamefont {K.-E.}\ \bibnamefont {Peiponen}}, \bibinfo {author} {\bibfnamefont {J.~J.}\ \bibnamefont {Saarinen}},\ and\ \bibinfo {author} {\bibfnamefont {E.~M.}\ \bibnamefont {Vartiainen}},\ }\href@noop {} {\emph {\bibinfo {title} {Kramers-Kronig relations in optical materials research}}}\ (\bibinfo  {publisher} {Springer},\ \bibinfo {year} {2005})\BibitemShut {NoStop}%
\bibitem [{\citenamefont {Jackson}(1998)}]{jackson1998appendix}%
  \BibitemOpen
  \bibfield  {author} {\bibinfo {author} {\bibfnamefont {J.~D.}\ \bibnamefont {Jackson}},\ }\bibinfo {title} {Appendix on units and dimensions},\ in\ \href@noop {} {\emph {\bibinfo {booktitle} {Classical Electrodynamics}}}\ (\bibinfo  {publisher} {John Wiley \& Sons},\ \bibinfo {year} {1998})\ pp.\ \bibinfo {pages} {775--784},\ \bibinfo {edition} {3rd}\ ed.\BibitemShut {Stop}%
\bibitem [{Note3()}]{Note3}%
  \BibitemOpen
  \bibinfo {note} {Conversion to the SI system of electromagnetic quantities and units is achieved via the relation $\chi ^{SI}= 4\pi \times 10^{-2} \chi ^{Gaussian}$}\BibitemShut {NoStop}%
\bibitem [{\citenamefont {Komissarov}\ \emph {et~al.}(2024)\citenamefont {Komissarov}, \citenamefont {Holder},\ and\ \citenamefont {Queiroz}}]{komissarov_quantum_2024}%
  \BibitemOpen
  \bibfield  {author} {\bibinfo {author} {\bibfnamefont {I.}~\bibnamefont {Komissarov}}, \bibinfo {author} {\bibfnamefont {T.}~\bibnamefont {Holder}},\ and\ \bibinfo {author} {\bibfnamefont {R.}~\bibnamefont {Queiroz}},\ }\bibfield  {title} {\bibinfo {title} {The quantum geometric origin of capacitance in insulators},\ }\href {https://doi.org/10.1038/s41467-024-48808-x} {\bibfield  {journal} {\bibinfo  {journal} {Nature Communications}\ }\textbf {\bibinfo {volume} {15}},\ \bibinfo {pages} {4621} (\bibinfo {year} {2024})}\BibitemShut {NoStop}%
\bibitem [{\citenamefont {Tomarchio}\ \emph {et~al.}(2022)\citenamefont {Tomarchio}, \citenamefont {Mosesso}, \citenamefont {Macis}, \citenamefont {Grilli}, \citenamefont {Romani}, \citenamefont {Guidi}, \citenamefont {Zhu}, \citenamefont {Feng}, \citenamefont {Zacchigna}, \citenamefont {Petrarca} \emph {et~al.}}]{tomarchio2022electrodynamics}%
  \BibitemOpen
  \bibfield  {author} {\bibinfo {author} {\bibfnamefont {L.}~\bibnamefont {Tomarchio}}, \bibinfo {author} {\bibfnamefont {L.}~\bibnamefont {Mosesso}}, \bibinfo {author} {\bibfnamefont {S.}~\bibnamefont {Macis}}, \bibinfo {author} {\bibfnamefont {A.}~\bibnamefont {Grilli}}, \bibinfo {author} {\bibfnamefont {M.}~\bibnamefont {Romani}}, \bibinfo {author} {\bibfnamefont {M.~C.}\ \bibnamefont {Guidi}}, \bibinfo {author} {\bibfnamefont {K.}~\bibnamefont {Zhu}}, \bibinfo {author} {\bibfnamefont {X.}~\bibnamefont {Feng}}, \bibinfo {author} {\bibfnamefont {M.}~\bibnamefont {Zacchigna}}, \bibinfo {author} {\bibfnamefont {M.}~\bibnamefont {Petrarca}}, \emph {et~al.},\ }\bibfield  {title} {\bibinfo {title} {Electrodynamics of {MnBi$_2$Te$_4$} intrinsic magnetic topological insulators},\ }\href {https://doi.org/10.1038/s41427-022-00429-w} {\bibfield  {journal} {\bibinfo  {journal} {NPG Asia Materials}\ }\textbf {\bibinfo {volume} {14}},\ \bibinfo {pages} {82} (\bibinfo {year} {2022})}\BibitemShut {NoStop}%
\bibitem [{\citenamefont {Fu}\ \emph {et~al.}(2020)\citenamefont {Fu}, \citenamefont {Liu},\ and\ \citenamefont {Yan}}]{doi:10.1126/sciadv.aaz0948}%
  \BibitemOpen
  \bibfield  {author} {\bibinfo {author} {\bibfnamefont {H.}~\bibnamefont {Fu}}, \bibinfo {author} {\bibfnamefont {C.-X.}\ \bibnamefont {Liu}},\ and\ \bibinfo {author} {\bibfnamefont {B.}~\bibnamefont {Yan}},\ }\bibfield  {title} {\bibinfo {title} {Exchange bias and quantum anomalous hall effect in the {MnBi\textsubscript{2}Te\textsubscript{4}}/{CrI\textsubscript{3}} heterostructure},\ }\href {https://doi.org/10.1126/sciadv.aaz0948} {\bibfield  {journal} {\bibinfo  {journal} {Science Advances}\ }\textbf {\bibinfo {volume} {6}},\ \bibinfo {pages} {eaaz0948} (\bibinfo {year} {2020})},\ \Eprint {https://arxiv.org/abs/https://www.science.org/doi/pdf/10.1126/sciadv.aaz0948} {https://www.science.org/doi/pdf/10.1126/sciadv.aaz0948} \BibitemShut {NoStop}%
\bibitem [{\citenamefont {Mostofi}\ \emph {et~al.}(2014)\citenamefont {Mostofi}, \citenamefont {Yates}, \citenamefont {Pizzi}, \citenamefont {Lee}, \citenamefont {Souza}, \citenamefont {Vanderbilt},\ and\ \citenamefont {Marzari}}]{MOSTOFI20142309}%
  \BibitemOpen
  \bibfield  {author} {\bibinfo {author} {\bibfnamefont {A.~A.}\ \bibnamefont {Mostofi}}, \bibinfo {author} {\bibfnamefont {J.~R.}\ \bibnamefont {Yates}}, \bibinfo {author} {\bibfnamefont {G.}~\bibnamefont {Pizzi}}, \bibinfo {author} {\bibfnamefont {Y.-S.}\ \bibnamefont {Lee}}, \bibinfo {author} {\bibfnamefont {I.}~\bibnamefont {Souza}}, \bibinfo {author} {\bibfnamefont {D.}~\bibnamefont {Vanderbilt}},\ and\ \bibinfo {author} {\bibfnamefont {N.}~\bibnamefont {Marzari}},\ }\bibfield  {title} {\bibinfo {title} {An updated version of wannier90: A tool for obtaining maximally-localised wannier functions},\ }\href {https://doi.org/https://doi.org/10.1016/j.cpc.2014.05.003} {\bibfield  {journal} {\bibinfo  {journal} {Computer Physics Communications}\ }\textbf {\bibinfo {volume} {185}},\ \bibinfo {pages} {2309} (\bibinfo {year} {2014})}\BibitemShut {NoStop}%
\bibitem [{\citenamefont {Pizzi}\ \emph {et~al.}(2020)\citenamefont {Pizzi}, \citenamefont {Vitale}, \citenamefont {Arita}, \citenamefont {Blügel}, \citenamefont {Freimuth}, \citenamefont {Géranton}, \citenamefont {Gibertini}, \citenamefont {Gresch}, \citenamefont {Johnson}, \citenamefont {Koretsune}, \citenamefont {Ibañez-Azpiroz}, \citenamefont {Lee}, \citenamefont {Lihm}, \citenamefont {Marchand}, \citenamefont {Marrazzo}, \citenamefont {Mokrousov}, \citenamefont {Mustafa}, \citenamefont {Nohara}, \citenamefont {Nomura}, \citenamefont {Paulatto}, \citenamefont {Poncé}, \citenamefont {Ponweiser}, \citenamefont {Qiao}, \citenamefont {Thöle}, \citenamefont {Tsirkin}, \citenamefont {Wierzbowska}, \citenamefont {Marzari}, \citenamefont {Vanderbilt}, \citenamefont {Souza}, \citenamefont {Mostofi},\ and\ \citenamefont {Yates}}]{Pizzi_2020}%
  \BibitemOpen
  \bibfield  {author} {\bibinfo {author} {\bibfnamefont {G.}~\bibnamefont {Pizzi}}, \bibinfo {author} {\bibfnamefont {V.}~\bibnamefont {Vitale}}, \bibinfo {author} {\bibfnamefont {R.}~\bibnamefont {Arita}}, \bibinfo {author} {\bibfnamefont {S.}~\bibnamefont {Blügel}}, \bibinfo {author} {\bibfnamefont {F.}~\bibnamefont {Freimuth}}, \bibinfo {author} {\bibfnamefont {G.}~\bibnamefont {Géranton}}, \bibinfo {author} {\bibfnamefont {M.}~\bibnamefont {Gibertini}}, \bibinfo {author} {\bibfnamefont {D.}~\bibnamefont {Gresch}}, \bibinfo {author} {\bibfnamefont {C.}~\bibnamefont {Johnson}}, \bibinfo {author} {\bibfnamefont {T.}~\bibnamefont {Koretsune}}, \bibinfo {author} {\bibfnamefont {J.}~\bibnamefont {Ibañez-Azpiroz}}, \bibinfo {author} {\bibfnamefont {H.}~\bibnamefont {Lee}}, \bibinfo {author} {\bibfnamefont {J.-M.}\ \bibnamefont {Lihm}}, \bibinfo {author} {\bibfnamefont {D.}~\bibnamefont {Marchand}}, \bibinfo {author} {\bibfnamefont {A.}~\bibnamefont {Marrazzo}}, \bibinfo {author} {\bibfnamefont
  {Y.}~\bibnamefont {Mokrousov}}, \bibinfo {author} {\bibfnamefont {J.~I.}\ \bibnamefont {Mustafa}}, \bibinfo {author} {\bibfnamefont {Y.}~\bibnamefont {Nohara}}, \bibinfo {author} {\bibfnamefont {Y.}~\bibnamefont {Nomura}}, \bibinfo {author} {\bibfnamefont {L.}~\bibnamefont {Paulatto}}, \bibinfo {author} {\bibfnamefont {S.}~\bibnamefont {Poncé}}, \bibinfo {author} {\bibfnamefont {T.}~\bibnamefont {Ponweiser}}, \bibinfo {author} {\bibfnamefont {J.}~\bibnamefont {Qiao}}, \bibinfo {author} {\bibfnamefont {F.}~\bibnamefont {Thöle}}, \bibinfo {author} {\bibfnamefont {S.~S.}\ \bibnamefont {Tsirkin}}, \bibinfo {author} {\bibfnamefont {M.}~\bibnamefont {Wierzbowska}}, \bibinfo {author} {\bibfnamefont {N.}~\bibnamefont {Marzari}}, \bibinfo {author} {\bibfnamefont {D.}~\bibnamefont {Vanderbilt}}, \bibinfo {author} {\bibfnamefont {I.}~\bibnamefont {Souza}}, \bibinfo {author} {\bibfnamefont {A.~A.}\ \bibnamefont {Mostofi}},\ and\ \bibinfo {author} {\bibfnamefont {J.~R.}\ \bibnamefont {Yates}},\ }\bibfield  {title}
  {\bibinfo {title} {Wannier90 as a community code: new features and applications},\ }\href {https://doi.org/10.1088/1361-648X/ab51ff} {\bibfield  {journal} {\bibinfo  {journal} {Journal of Physics: Condensed Matter}\ }\textbf {\bibinfo {volume} {32}},\ \bibinfo {pages} {165902} (\bibinfo {year} {2020})}\BibitemShut {NoStop}%
\bibitem [{\citenamefont {Tsirkin}(2021)}]{tsirkin_high_2021}%
  \BibitemOpen
  \bibfield  {author} {\bibinfo {author} {\bibfnamefont {S.~S.}\ \bibnamefont {Tsirkin}},\ }\bibfield  {title} {\bibinfo {title} {High performance {Wannier} interpolation of {Berry} curvature and related quantities with {WannierBerri} code},\ }\bibfield  {journal} {\bibinfo  {journal} {npj Computational Materials}\ }\textbf {\bibinfo {volume} {7}},\ \href {https://doi.org/10.1038/s41524-021-00498-5} {10.1038/s41524-021-00498-5} (\bibinfo {year} {2021})\BibitemShut {NoStop}%
\bibitem [{\citenamefont {Varnava}\ and\ \citenamefont {Vanderbilt}(2018)}]{PhysRevB.98.245117}%
  \BibitemOpen
  \bibfield  {author} {\bibinfo {author} {\bibfnamefont {N.}~\bibnamefont {Varnava}}\ and\ \bibinfo {author} {\bibfnamefont {D.}~\bibnamefont {Vanderbilt}},\ }\bibfield  {title} {\bibinfo {title} {Surfaces of axion insulators},\ }\href {https://doi.org/10.1103/PhysRevB.98.245117} {\bibfield  {journal} {\bibinfo  {journal} {Phys. Rev. B}\ }\textbf {\bibinfo {volume} {98}},\ \bibinfo {pages} {245117} (\bibinfo {year} {2018})}\BibitemShut {NoStop}%
\bibitem [{\citenamefont {Bianco}\ and\ \citenamefont {Resta}(2011)}]{PhysRevB.84.241106}%
  \BibitemOpen
  \bibfield  {author} {\bibinfo {author} {\bibfnamefont {R.}~\bibnamefont {Bianco}}\ and\ \bibinfo {author} {\bibfnamefont {R.}~\bibnamefont {Resta}},\ }\bibfield  {title} {\bibinfo {title} {Mapping topological order in coordinate space},\ }\href {https://doi.org/10.1103/PhysRevB.84.241106} {\bibfield  {journal} {\bibinfo  {journal} {Phys. Rev. B}\ }\textbf {\bibinfo {volume} {84}},\ \bibinfo {pages} {241106(R)} (\bibinfo {year} {2011})}\BibitemShut {NoStop}%
\bibitem [{\citenamefont {Marrazzo}\ and\ \citenamefont {Resta}(2017)}]{PhysRevB.95.121114}%
  \BibitemOpen
  \bibfield  {author} {\bibinfo {author} {\bibfnamefont {A.}~\bibnamefont {Marrazzo}}\ and\ \bibinfo {author} {\bibfnamefont {R.}~\bibnamefont {Resta}},\ }\bibfield  {title} {\bibinfo {title} {Locality of the anomalous hall conductivity},\ }\href {https://doi.org/10.1103/PhysRevB.95.121114} {\bibfield  {journal} {\bibinfo  {journal} {Phys. Rev. B}\ }\textbf {\bibinfo {volume} {95}},\ \bibinfo {pages} {121114(R)} (\bibinfo {year} {2017})}\BibitemShut {NoStop}%
\bibitem [{\citenamefont {Li}\ \emph {et~al.}(2025{\natexlab{b}})\citenamefont {Li}, \citenamefont {Wu},\ and\ \citenamefont {Weng}}]{li2025high}%
  \BibitemOpen
  \bibfield  {author} {\bibinfo {author} {\bibfnamefont {J.}~\bibnamefont {Li}}, \bibinfo {author} {\bibfnamefont {Q.}~\bibnamefont {Wu}},\ and\ \bibinfo {author} {\bibfnamefont {H.}~\bibnamefont {Weng}},\ }\bibfield  {title} {\bibinfo {title} {High-chern-number quantum anomalous hall insulators in mixing-stacked mnbi2te4 thin films},\ }\href {https://doi.org/https://doi.org/10.1038/s41535-025-00775-2} {\bibfield  {journal} {\bibinfo  {journal} {npj Quantum Materials}\ }\textbf {\bibinfo {volume} {10}},\ \bibinfo {pages} {53} (\bibinfo {year} {2025}{\natexlab{b}})}\BibitemShut {NoStop}%
\end{thebibliography}

\end{document}